\documentclass[10pt, conference]{IEEEtran}

\let\labelindent\relax
\usepackage{enumitem}

\usepackage{amsmath}
\usepackage{graphicx}
\usepackage{multirow}
\usepackage{subfig}
\usepackage[linesnumbered, ruled]{algorithm2e}
\usepackage{url}
\begin{document}

\title{\textit{Multi-Tenant Virtual GPUs for Optimising Performance of a Financial Risk Application}}

%\author{
%\IEEEauthorblockN{Blesson Varghese}
%\IEEEauthorblockA{Queen's University Belfast, UK\\
%Email: varghese@qub.ac.uk}
%\and
%\IEEEauthorblockN{Javier Prades, Carlos Rea\~{n}o and Federico Silla}
%\IEEEauthorblockA{Universitat Polit\`{e}cnica de Val\`{e}ncia, Spain\\
%Email: \{japraga,carregon\}@gap.upv.es, fsilla@disca.upv.es}
%}

\author{
	\IEEEauthorblockN{Javier Prades}
	\IEEEauthorblockA  {
		Universitat Polit\`{e}cnica de Val\`{e}ncia, Spain\\japraga@gap.upv.es
	}
	\and
	\IEEEauthorblockN{Blesson Varghese}
	\IEEEauthorblockA  {
		Queen's University Belfast, UK\\varghese@qub.ac.uk
	}
	\and
	\IEEEauthorblockN{Carlos Rea\~{n}o and Federico Silla}
	\IEEEauthorblockA  {
		Universitat Polit\`{e}cnica de Val\`{e}ncia, Spain\\carregon@gap.upv.es, fsilla@disca.upv.es
	}
%	\and
%	\IEEEauthorblockN{Federico Silla}
%	\IEEEauthorblockA  {
%		Universitat Polit\`{e}cnica de Val\`{e}ncia, Spain\\fsilla@disca.upv.es
%	}	
	
}

\maketitle
\thispagestyle{plain}
\pagestyle{plain}

%Abstract
\noindent Graphics Processing Units (GPUs) are becoming popular accelerators in modern High-Performance Computing (HPC) clusters. Installing GPUs on each node of the cluster is not efficient resulting in high costs and power consumption as well as underutilisation of the accelerator. The research reported in this paper is motivated towards the use of few physical GPUs by providing cluster nodes access to remote GPUs on-demand for a financial risk application. We hypothesise that sharing GPUs between several nodes, referred to as multi-tenancy, reduces the execution time and energy consumed by an application. Two data transfer modes between the CPU and the GPUs, namely concurrent and sequential, are explored. The key result from the experiments is that multi-tenancy with few physical GPUs using sequential data transfers lowers the execution time and the energy consumed, thereby improving the overall performance of the application.

%Keywords
%\input{keywords}

\IEEEpeerreviewmaketitle

\section{Introduction}
\label{introduction}
Hardware accelerators are achieving a prominent role in modern High-Performance Computing (HPC) clusters for making applications faster. This is evidenced by four out of top ten supercomputers listed on Top500 (http://top500.org) and the top ten supercomputers listed on Green500 (http://www.green500.org) in November 2015 have employed hardware accelerators, such as Graphics Processing Units (GPU). Incorporating GPUs in large clusters allows for heterogeneity, thus making it possible for an application to exploit the regular processor as well as the accelerator \cite{hetcluster-1,hetcluster-2}. 

Clusters can now be set up to employ a small number of GPUs by providing applications shared access to remote GPUs on-demand \cite{multitenancy-1, multitenancy-2}. Such a set up is feasible on a limited budget because not only are a few GPUs used to provide acceleration, but also the energy consumed is well justified since the GPUs are well utilised in the cluster \cite{gpu-power1, rcuda-energy}. This is possible as a result of maturing GPU virtualisation technologies that facilitate virtual GPUs (vGPUs) in a cluster. An application can request Acceleration-as-a-Service\cite{AaaS-1} from one or many vGPUs. One vGPU can reside on a physical GPU (pGPU), referred to as \textit{single tenancy}, but is limiting in that multiple applications cannot make use of the same pGPU since it is exclusively locked for a single application. When multiple vGPUs reside on the same pGPU, otherwise known as \textit{multi-tenancy}, either the same application has access to a pool of vGPUs on the same pGPU or multiple applications can share the same pGPU. We hypothesise that using multi-tenancy can improve the performance of an application.
 
\begin{figure}[t]
	\centering
		\includegraphics[width=0.5\textwidth]{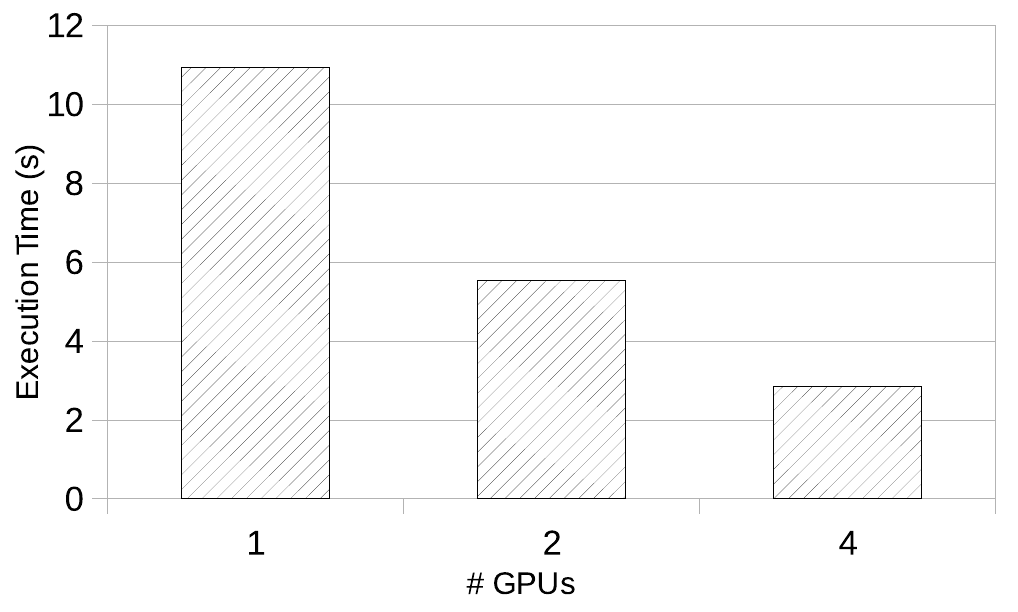}
		\caption{Execution time of the financial application on multiple local GPUs}
		\label{cuda_performance}
\end{figure}

Numerous challenges arise when multiple GPUs are shared across a cluster for an application, of which three are considered in this paper. The challenges are addressed in this paper by exploring remote CUDA (rCUDA) \cite{PARCOToni}, a GPU virtualisation framework, for improving the performance of a real-world case study employed in the financial industry. The application typically runs in a cluster environment, but can hugely benefit from GPU acceleration for deriving important risk metrics in real-time. The benefit of executing the application on multiple physical GPUs is shown in Figure~\ref{cuda_performance}. 
We hypothesise that using a large number of vGPUs can further optimise application performance. However, the following three challenges and research questions arise, which are addressed in this paper:
(i) Data will need to be transferred from the CPU to the vGPUs for computations. However, data transfer will be restricted by bottlenecks due to limited bandwidth which affects the overall scalability of the application. Hence, ``What data transfer approaches can mitigate the effect of data bottlenecks?''
(ii) Multi-tenancy may degrade application performance since the underlying hardware resource is shared. This results in increased execution time and consequently higher energy consumption. Hence, ``How can vGPUs be shared effectively to optimise application performance and energy consumed?''
(iii) Using multi-tenancy an application can be deployed in multiple ways. For example, an application can be executed on 2 vGPUs residing on 1 pGPU or 8 vGPUs residing on 1 pGPU. These possibilities significantly increase with multiple pGPUs. Each deployment option consumes different amounts of energy and impacts the overall execution time. Hence, ``Can performance and energy of an application be estimated in the multi-tenancy approach?''

To address the above challenges we propose two data transfer approaches, namely concurrent and sequential, for transferring data with the aim of mitigating the effect of data bottlenecks. In the context of the financial application, the sequential data transfer approach is expected to improve performance since data transfers from the CPU to the GPU and GPU computations can be overlapped for multiple pGPUs. The approach is further extended for overlapping the data movement and computation time for multiple vGPUs on the same pGPU resulting in a further improvement in performance of the application. The key result is that the financial application can be executed under two seconds for deriving risk metrics in an energy efficient manner on the same hardware compared to single tenancy thus confirming our initial hypothesis. Performance and energy consumed by the application are modelled to determine the combination of vGPUs on a pGPU that can maximise performance and GPU utilisation and at the same time minimise the energy consumed. 

The key contributions of this research are: (i) investigating the lack of scalability due to data transfer from CPU to the GPU in the context of the financial risk application, (ii) proposing two approaches to transfer data, namely concurrent and sequential, (iii) evaluating the above data transfer approaches in the context of single-tenancy for overlapping computations and data transfer of multiple pGPUs, (iv) developing an approach that exploits multi-tenancy for overlapping computations and data transfer of multiple virtual GPUs on the same physical GPU to optimise the performance of the application, (v) evaluating the performance of the application, considering execution time, GPU utilisation and energy consumed by the application, and (vi) developing a mathematical model to derive deployment options for the application by estimating performance and energy of different combinations of virtual GPUs mapped onto physical GPUs. 

The remainder of this paper is organised as follows. 
Section \ref{relatedwork} highlights related work in the area of HPC solutions for GPU virtualisation and financial risk applications. 
Section \ref{rcuda} briefly presents the rCUDA framework. 
Section \ref{application} considers a financial risk application for evaluating the feasibility of multi-tenancy for improving performance. 
Section \ref{evaluation} presents the platform, experiments performed and the key results obtained. 
Section \ref{conclusions} concludes this paper.

\section{Related Work}
\label{relatedwork}
High Performance Computing (HPC) solutions are exploited in the financial risk industry to accelerate the underlying computations of applications. This reduces overall execution times making such applications fit for real-time use. Solutions range from small scale clusters \cite{smallcluster1, smallcluster2} to large supercomputers \cite{supercomputer1, supercomputer2}. More recently, hardware accelerators with multi-core and many-core processors are employed. For example, financial risk applications are accelerated on Cell BE processors \cite{cellbe-1, cellbe-2}, FPGAs \cite{fpga3, fpga4} and GPUs \cite{gpu3, gpu5}. 

HPC clusters offering heterogeneous solutions by using hardware accelerators, such as GPUs, along with processors on nodes are feasible \cite{hetcluster-1, hetcluster-2}. Clusters can be set up to incorporate a GPU on each node. This is not an efficient solution for accelerating an application because of the relatively high cost of GPUs, high power consumption of nodes using GPUs and the under utilisation of GPUs (applications do not require acceleration of GPUs during their entire execution). However, a more efficient solution would be if nodes executing an application can access GPUs when required. This can be facilitated by GPU virtualisation. Currently there are no solutions available for the financial risk industry to harness the potential of GPU virtualisation. In this paper, we investigate the use of virtual GPUs for a financial risk application. 

The mechanism of GPU virtualisation allows nodes of a cluster that do not own a physical GPU for accelerating computations of applications that run on it to remotely access GPUs. Acceleration is obtained as a service seamlessly to a requesting node without being aware of accessing remote GPUs. A single application (running on a Virtual Machine (VM) or on a node of a cluster without a hardware accelerator) benefits from the acceleration of a remotely located single GPU or multiple GPUs to reduce execution time. The rate of GPU utilisation can be increased since multiple applications can access the same GPU. This in turn reduces the number of GPUs that need to be installed in a cluster, and reduces the cost spent on energy consumption, cooling, physical space and maintenance, usually referred to as the Total Cost of Ownership (TCO). Furthermore, the source code of an application usually does not need any modification to reap the benefits of virtual GPUs.

GPU virtualisation is usually applied at the high-level Application Programming Interface (API) of GPUs because low level protocols used to interact with accelerators are proprietary and, additionally, not publicly available. Hence, APIs such as CUDA~\cite{CUDA} or OpenCL~\cite{OpenCL} are used. In this paper we use CUDA (Compute Unified Device Architecture) for an application that is used in the financial risk industry.  

There are several remote GPU virtualization frameworks supporting CUDA. GridCuda~\cite{Liang11} supports CUDA 3.2, although it is not publicly available. vCUDA~\cite{Shi09} supports the CUDA 3.2 and implements an unspecified subset of the CUDA runtime API. The communication protocol between the node that executes the application and the remote GPU has a considerable overhead, because of the costs incurred during encoding and decoding, which results in a noticeable drop of overall performance. GViM~\cite{Gupta09} is based on CUDA 1.1 and does not implement the entire runtime API. Furthermore, GViM is designed to be used on VMs so that applications executed on them can access GPUs located in the real host; GViM does not support the access of GPUs in remote nodes. gVirtuS~\cite{Giunta10} supports CUDA 2.3 an again implements only a small portion of the runtime API. For example, in the case of the memory management module, it implements only 17 out of the 37 available functions. Although it is intended mainly to be used by VMs for accessing real GPUs located in the same node, it facilitates TCP/IP communications between clients and servers, thus allowing the access to GPUs located in other nodes. DS-CUDA~\cite{Oikawa12} supports CUDA 4.1 and includes specific communication support for InfiniBand Verbs, thus reducing the overhead of communications between the node executing the application and the node owning the GPU. However, DS-CUDA is limited in that it does not allow data transfers with pinned memory and supports maximum data transfer of 32 MB. 

The rCUDA framework~\cite{PARCOToni} is binary compatible with CUDA~6.5 and implements the entire CUDA Runtime and Driver APIs (with the exception of graphics functions). It provides support for the libraries included within CUDA, such as cuBLAS or cuFFT. In addition, a number of underlying interconnection technologies are supported by making use of a set of runtime-loadable, network-specific communication modules (currently TCP/IP and InfiniBand). Concurrent virtualization services are made available to remote clients simultaneously demanding GPU acceleration by managing an independent GPU context for each client. rCUDA performs better than other publicly available GPU virtualisation frameworks (considered in Section~\ref{rcuda}) and is therefore chosen for this research.

\section{rCUDA}
\label{rcuda}
\begin{figure}[b!]
\centering
	\includegraphics[width=0.5\textwidth]{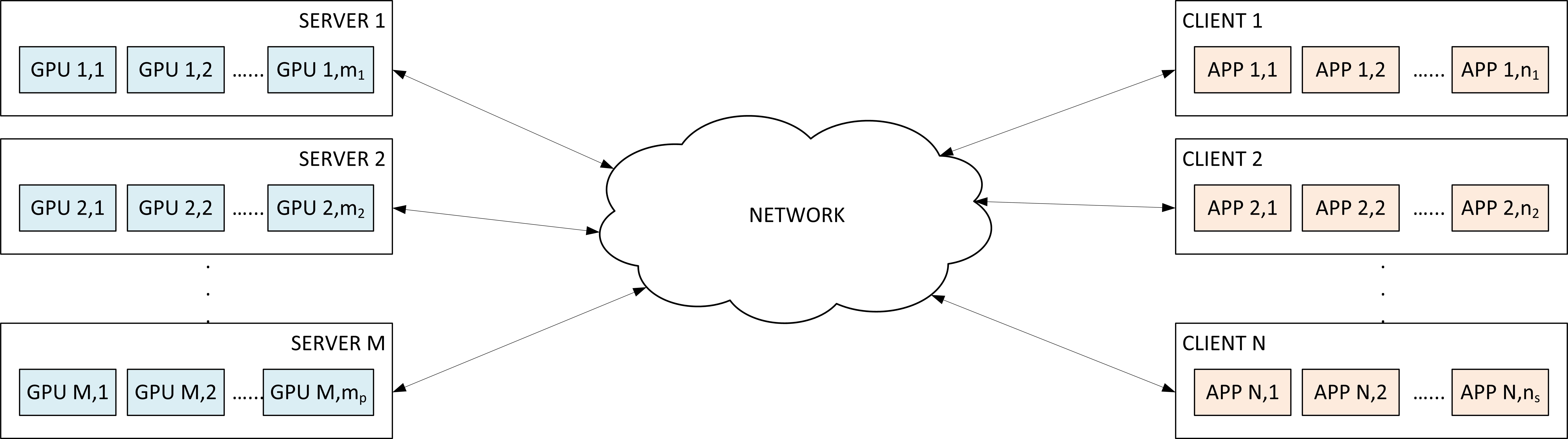}
	\caption{Distributed acceleration architecture facilitated by rCUDA}
	\label{figure2}
\end{figure}

The rCUDA framework, otherwise referred to as remote CUDA, is used in the research presented in this paper. As shown in
Figure~\ref{figure2}, the rCUDA framework is a client-server architecture. Numerous \textit{Clients} executing
applications that can benefit from hardware acceleration can concurrently access \textit{Servers} that have physical
GPUs on them. The client makes use of the remote GPU to accelerate part of the software code of the application,
referred to as kernel, running on it. The framework transparently handles the data management and the execution
management; the transfer of data between the local memory of the client, the local memory of the server and the GPU
memory, and the remote execution of the kernel.

\begin{figure}[ht]
\centering
	\includegraphics[width=0.5\textwidth]{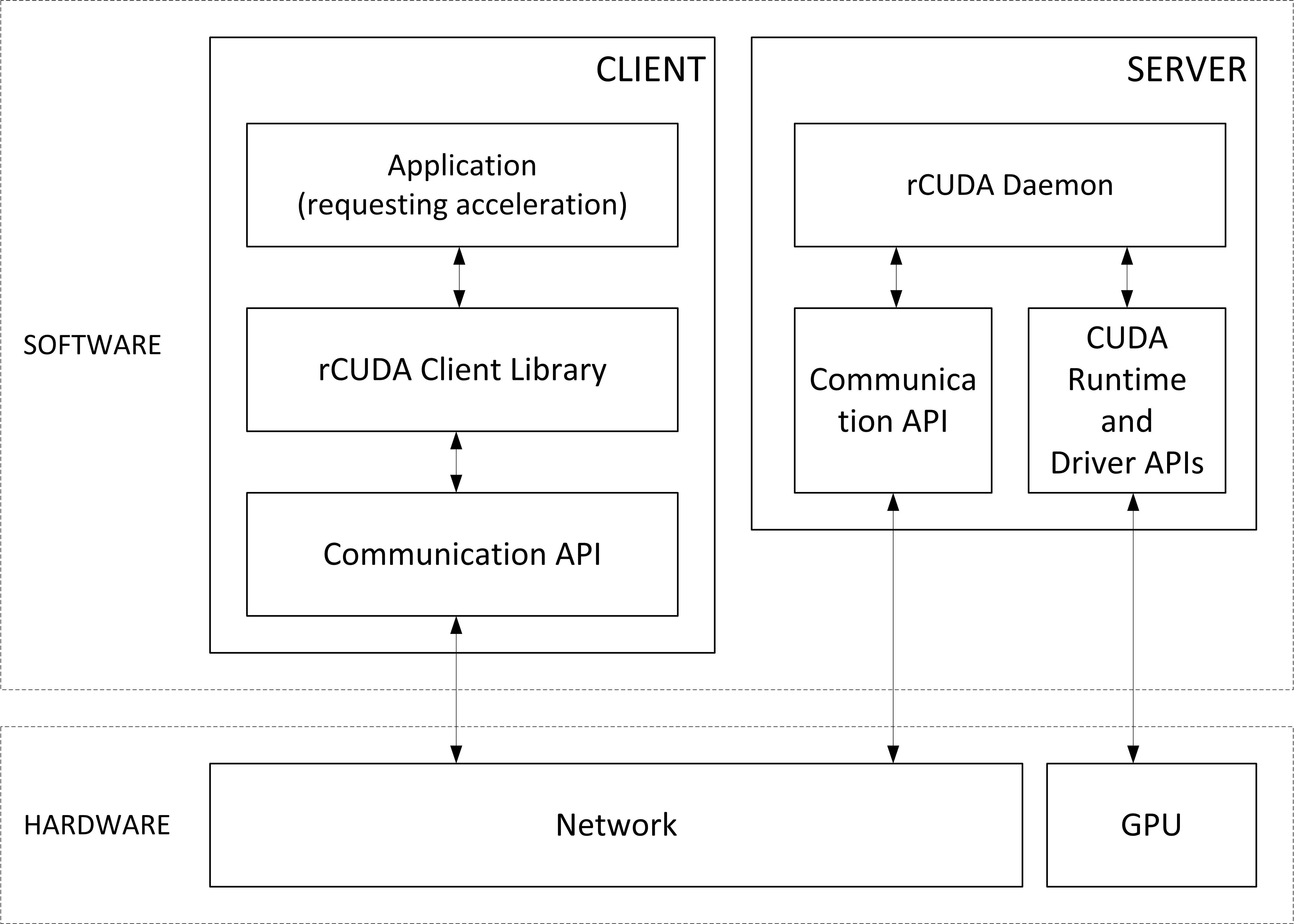}
	\caption{rCUDA client and server software/hardware stack}
	\label{figure3}
\end{figure}

Figure \ref{figure3} shows the hardware and software stack of the client and the rCUDA server. 
The client nodes that execute the application (shown in Figure~\ref{figure2}), make use of the rCUDA Client Library,
which is a wrapper around the CUDA Runtime and Driver APIs. The library is responsible for (i) intercepting calls made
by the application to a CUDA device, (ii) processing them for forwarding the calls to the remote rCUDA server, and (iii)
retrieving the results of the calls from the rCUDA server. 
On the other hand, each GPU server has an rCUDA daemon running on it which receives CUDA requests and executes them on the physical GPU.

An efficient communication protocol is developed for seamless execution between rCUDA clients and
servers. 
This protocol, using either regular TCP/IP sockets or the InfiniBand
Verbs API when this high performance interconnect is available in the cluster, is designed to provide lightweight
support to the remote CUDA operations provided by the external accelerator. The CUDA commands intercepted by the rCUDA
client wrapper are encapsulated into messages in the form of one or more packets that travel across the network towards
the rCUDA server. The format of the messages depends on the specific CUDA command transported. In general, the messages
have low network overheads. Every CUDA command forwarded to the remote GPU server is followed by a response message,
which acknowledges the success/failure of the operation requested on the remote server. 

\begin{figure}[t!]
\centering
	\includegraphics[width=0.48\textwidth]{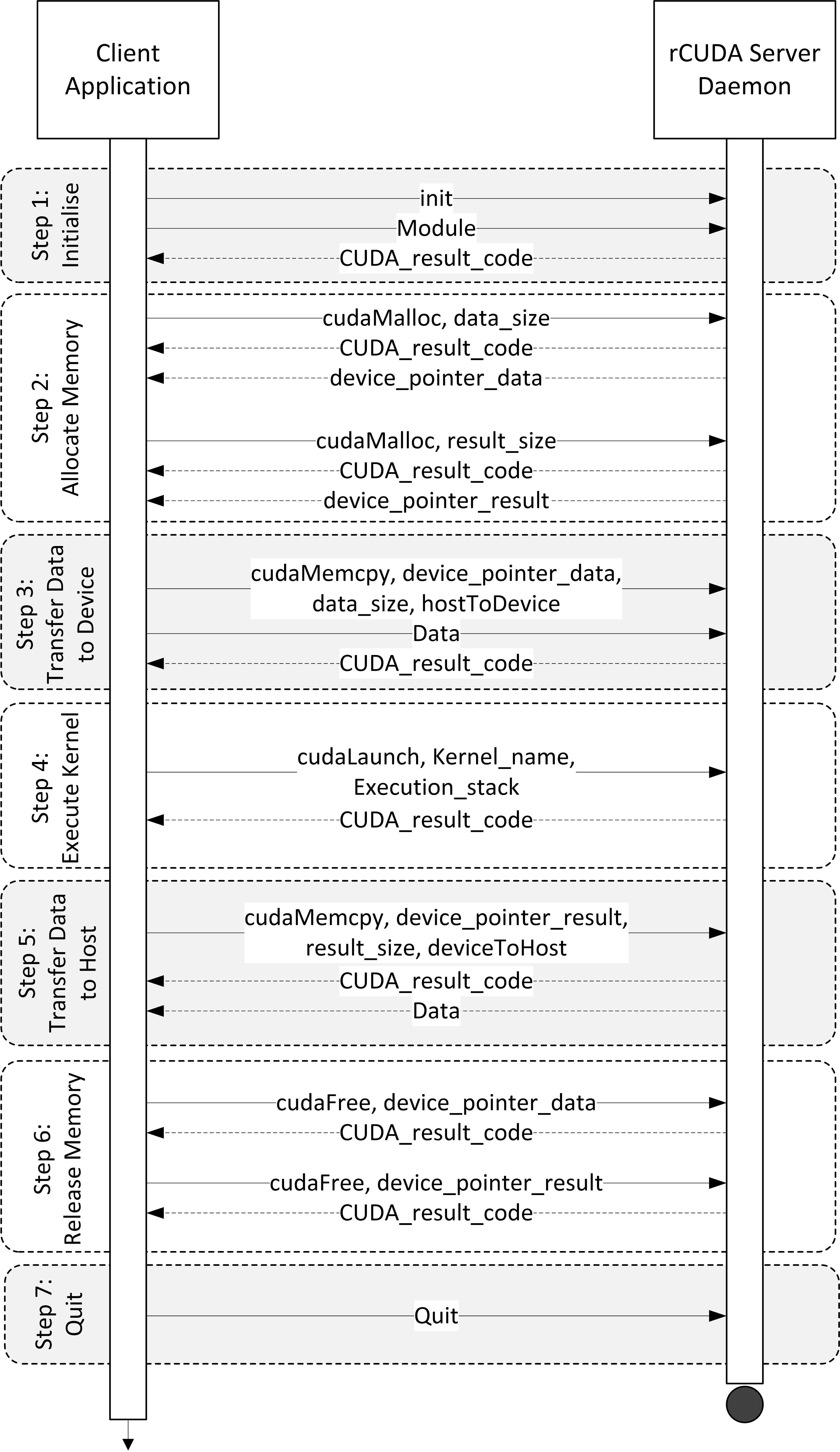}
	\caption{Communication sequence between a client and the rCUDA server daemon}
	\label{figure4}
\end{figure}

Figure \ref{figure4} shows an example of the communication between the rCUDA client and the rCUDA
daemon executing on the remote server. In this example, the following steps occur:

\begin{enumerate}[leftmargin=0pt]
\item[]\textit{Step 1 - Initialise}: The client establishes connection with the remote server automatically, and the request for acceleration services is intercepted and the GPU kernel along with related information such as statically allocated variables are sent to the server.
\item[]\textit{Step 2 - Allocate Memory}:
Based on the client request device memory is allocated on the GPU for data that will be required by the GPU kernel. The \texttt{cudaMalloc} requests are intercepted by the client and forwarded to the remote server. 
\item[]\textit{Step 3 - Transfer Data to Device}:
All data required by the kernel is transferred from the host to the remote device.
\item[]\textit{Step 4 - Execute Kernel}:
The GPU kernel is executed remotely on the rCUDA server.
\item[]\textit{Step 5 - Transfer Data to Host}:
After the execution of the kernel on the remote server the data is transmitted back to the host. 
\item[]\textit{Step 6 - Release Memory}:
The memory allocated on the remote device is released.
\item[]\textit{Step 7 - Quit}:
In this final step the client application stops communicating with the remote server. The rCUDA daemon executing on the server stops servicing the execution and releases the resources associated with the execution.  
\end{enumerate}

\begin{figure*}[t]
\centering
	\subfloat[Host pinned memory to device 
	memory]{\label{figure5a}\includegraphics[width=0.49\textwidth]{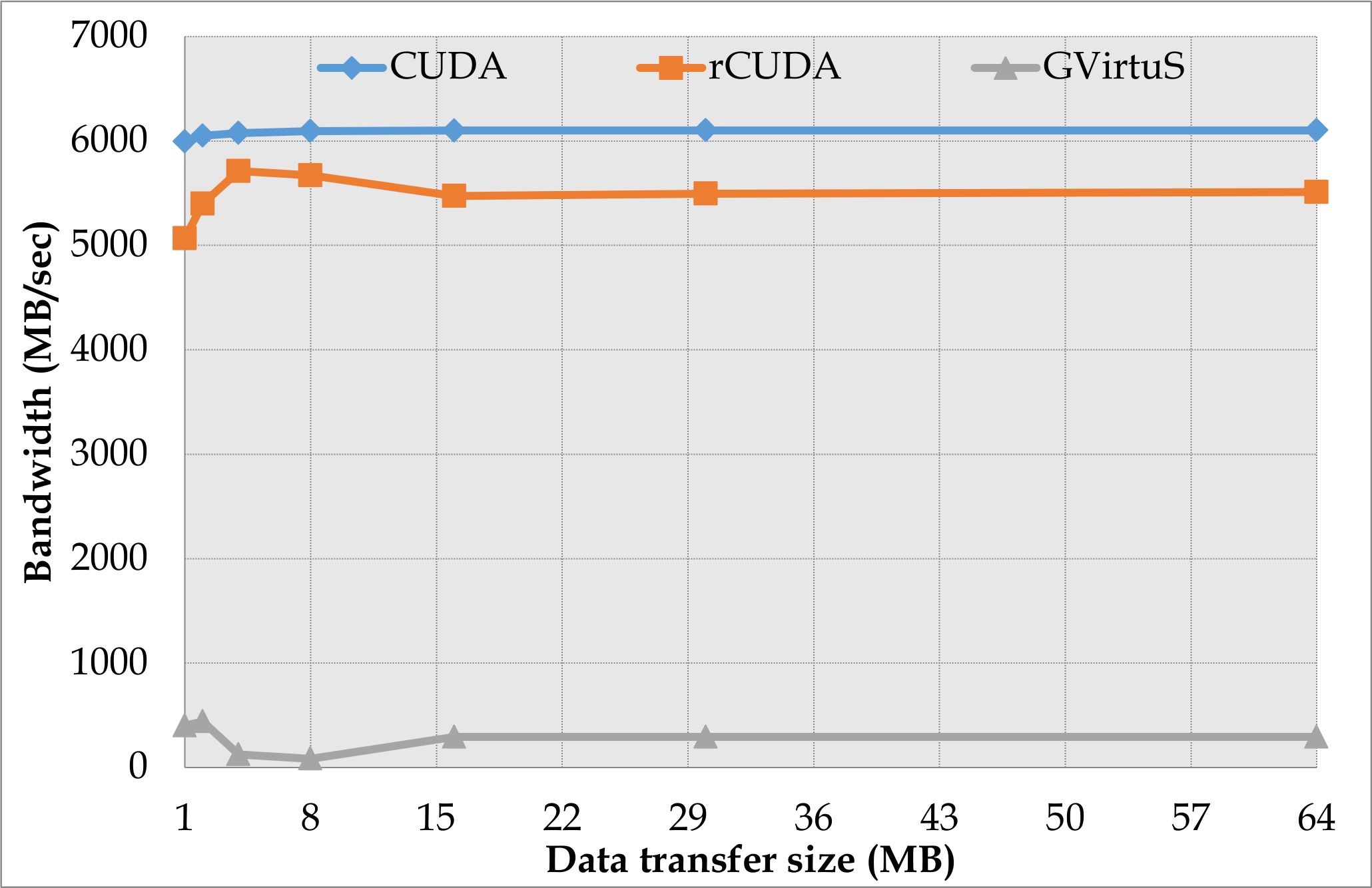}} \hfill
	\subfloat[Device memory to host pinned memory]{\label{figure5b}\includegraphics[width=0.49\textwidth]{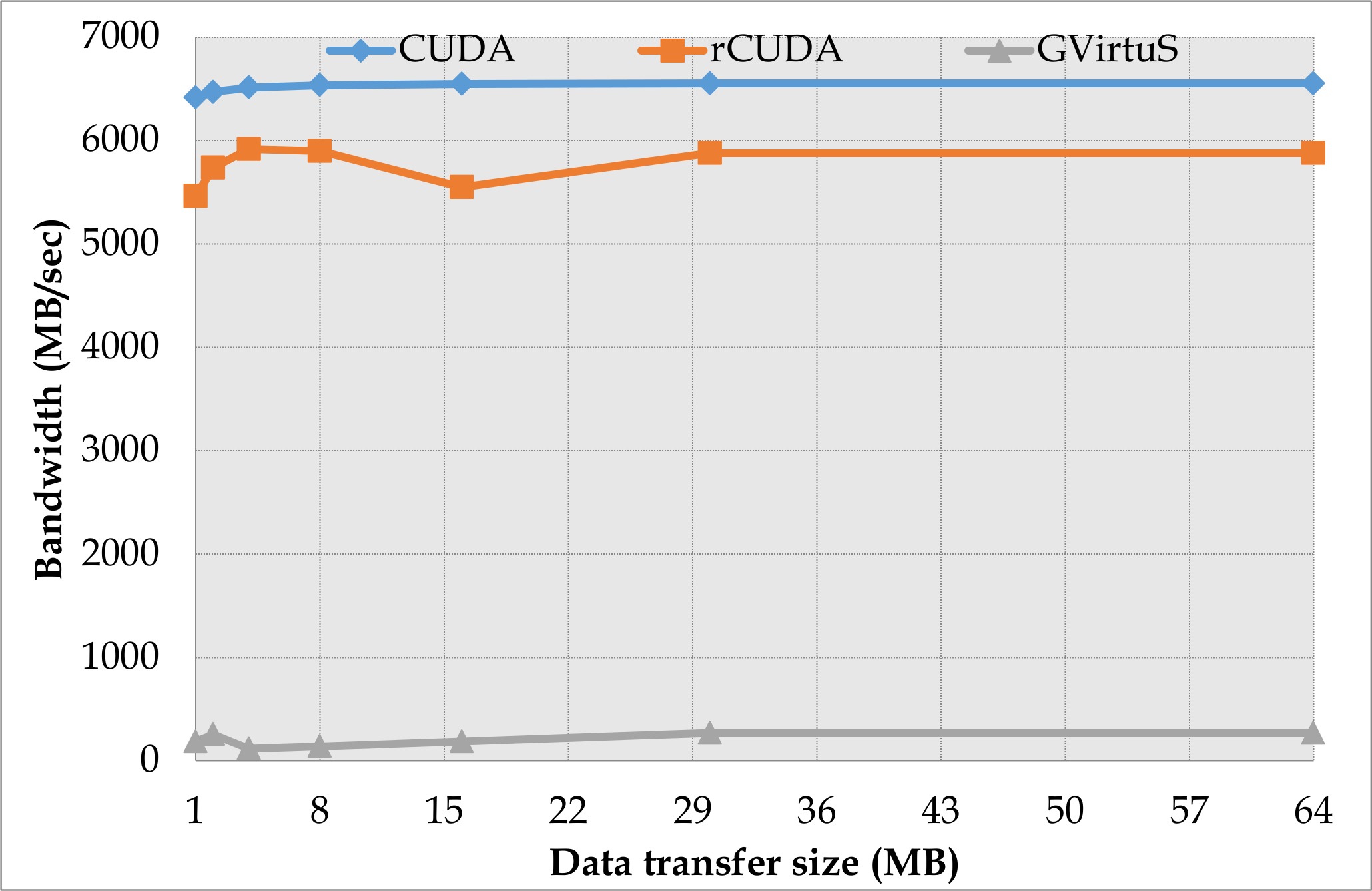}} \\
	\subfloat[Host pageable memory to device memory]{\label{figure5c}\includegraphics[width=0.49\textwidth]{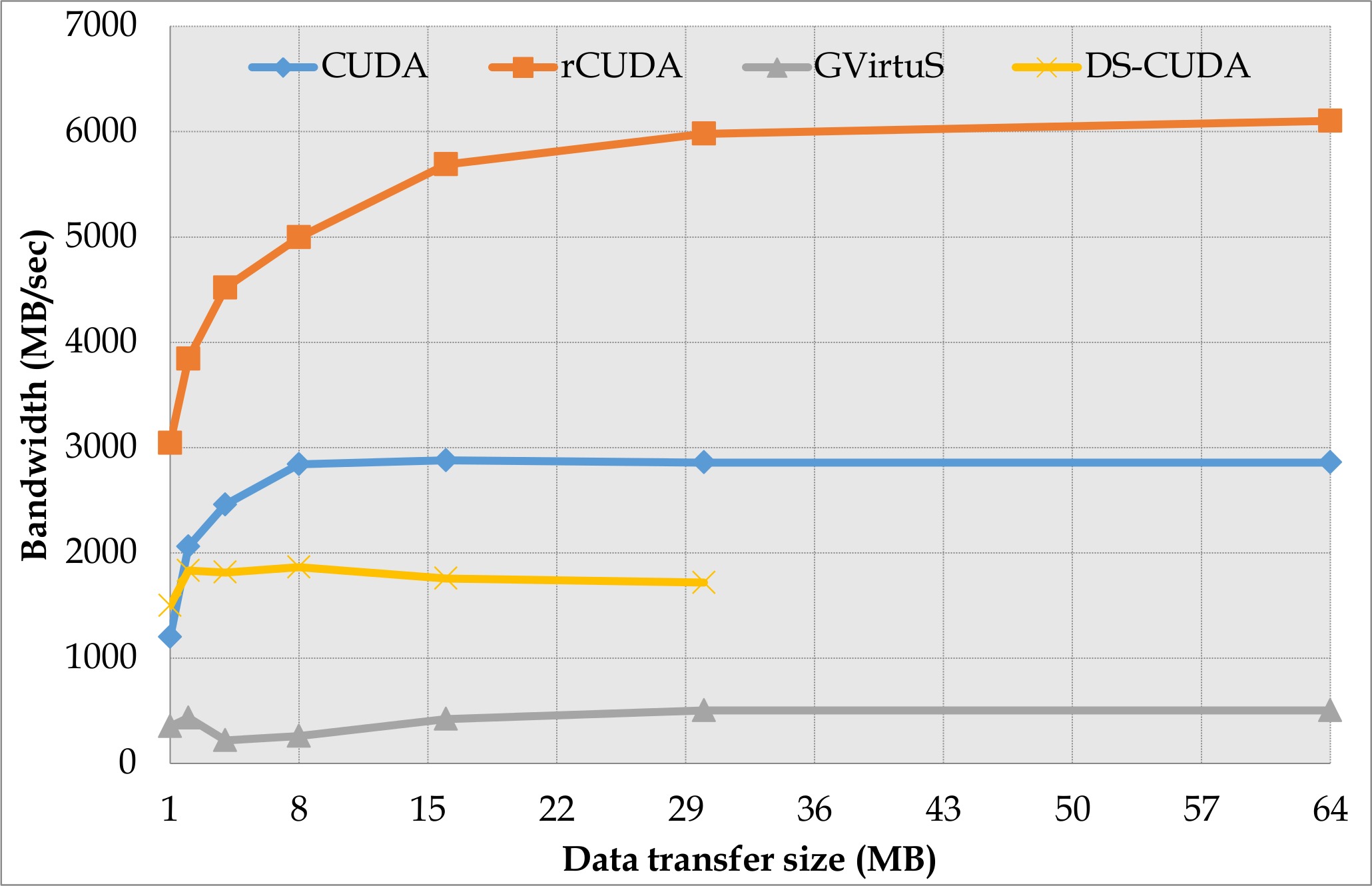}} \hfill
	\subfloat[Device memory to host pageable memory]{\label{figure5d}\includegraphics[width=0.49\textwidth]{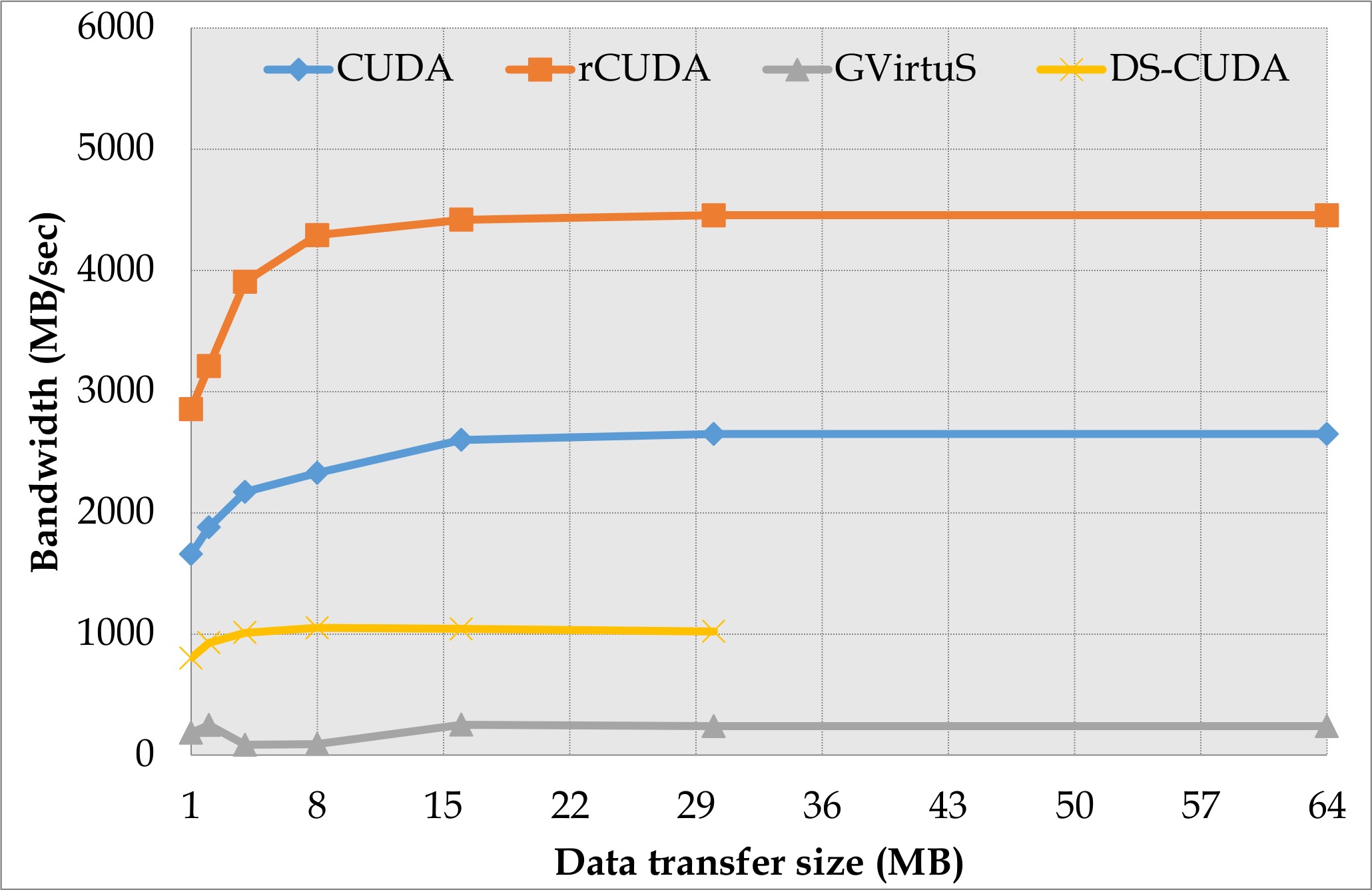}} \\
		
\caption{Comparison of bandwidth for pinned memory and pageable memory of rCUDA, DS-CUDA and gVirtuS using CUDA as a baseline reference (DS-CUDA does not support pinned memory)}
\label{figure5}
\end{figure*}

Figure~\ref{figure5} compares the performance of publicly available GPU virtualisation frameworks, namely DS-CUDA, gVirtuS and rCUDA by  using the
{\texttt{bandwidthTest}} benchmark from the NVIDIA CUDA Samples~\cite{NVIDIA_SDK}. Our choice of selecting rCUDA for this research is based on its superior performance over other frameworks as shown in the figure. The performance of CUDA 6.5 is used as the baseline reference. Bandwidth is used as a measure for comparing performance since it is a limiting factor for data transfers between host (CPU) memory and device (GPU) memory (data size can be in the order of MB) and affects the performance of the virtualisation frameworks. Other metrics such as latency are less relevant in this context.

The test-bed employed for carrying out the bandwidth performance experiments is presented later in Section~\ref{platform}. Virtual Machine (VMs) were not employed to simplify the experiments. The bandwidth test was run on a native domain and the server side of the virtualisation framework used was executed in a remote node. The InfiniBand FDR network technology was used to connect both nodes. The rCUDA and DS-CUDA frameworks made use of the InfiniBand Verbs API and gVirtuS made use of TCP/IP over InfiniBand since it cannot take advantage of the InfiniBand Verbs API.

The three virtualisation frameworks support different versions of CUDA which had to be used for obtaining the bandwidth benchmarks. DS-CUDA is compatible with CUDA 4.1, gVirtuS supports CUDA 2.3 and rCUDA supports CUDA 6.5. In our experience, employing different CUDA versions has minimal impact on bandwidth performance and therefore no additional noise was introduced by using different versions.

The following observations are made from Figure~\ref{figure4}. Firstly, CUDA achieves highest performance when pinned memory is used (refer Figure~\ref{figure5a} and Figure~\ref{figure5b}), achieving nearly a bandwidth of 6000~MB/s. The bandwidth is however reduced for copies using pageable memory (refer Figure~\ref{figure5c} and Figure~\ref{figure5d}). 

Secondly, Figure~\ref{figure5} shows that rCUDA outperforms DS-CUDA and gVirtuS. For copies using pageable memory rCUDA
even performs better than CUDA; this has been previously reported, which is due to the use of an efficient pipelined
communication between rCUDA clients and servers based on the use of internal and pre-allocated pinned memory buffers
~\cite{PARCOToni}. rCUDA and DS-CUDA support InfiniBand Verbs API and therefore have access to large bandwidths which
are available on this interconnect. However, DS-CUDA has relatively poor performance when compared to rCUDA. Therefore,
it is assumed that both frameworks manage the InfiniBand interconnect differently. DS-CUDA neither supports memory
copies larger than 32MB nor pinned memory. The performance of gVirtuS is significantly lower than the other frameworks.
It may be immediately inferred that this is because TCP/IP is used and has a lower bandwidth in comparison to InfiniBand
Verbs. However, using the \texttt{iperf} tool \cite{iperf}, TCP/IP over InfiniBand FDR provides approximately 1190 MB/s,
which is a noticeably larger bandwidth than the one achieved by gVirtuS. Therefore, the poor performance of gVirtuS may
be due to the inefficient handling of communication.

\section{Financial Risk Application}
\label{application}

A candidate application that can benefit from Acceleration-as-a-Service (AaaS) 
in HPC clusters is investigated in this section. We present such an application 
employed in the financial risk industry, referred to as \textit{`Aggregate Risk 
Analysis'}~\cite{Varghese2012} for validating the feasibility of our proposed multi-tenancy approach. The 
analysis of financial risk is underpinned by a simulation that is 
computationally intensive. Typically, this analysis is periodically performed on 
a routine basis on production clusters to derive important risk metrics. Such a 
set up is sufficient when the analysis does not need to be performed outside 
routine. 

Risk metrics will need to be obtained in real-time, such as in an online pricing scenario, in addition to routine executions. In such settings, a number of input parameters to the analysis will need to be varied to satisfy the customer. This generates a large number of requests to execute the analysis multiple times based on the complexity of the client's portfolio. It may not be feasible to furnish all these requests generated by single or multiple clients; it will be impossible to quickly obtain a large set of resources on an in-house cluster already provisioned for executing other routine jobs. Here, GPUs can play an important role in furnishing a large number of requests.

While GPUs can provide a feasible solution, employing a large number of GPUs to 
furnish bursts of requests will be expensive. As considered in 
Section~\ref{introduction} virtual GPUs are pragmatic and cost effective to 
minimise under utilisation. In this context, we leverage the acceleration 
offered by virtual GPUs in an HPC cluster to develop a faster application fit for 
use in real-time settings. The rCUDA framework suits such an application because 
minimal changes need to be brought about to the production cluster and the 
acceleration required for the analysis is obtained as a service from a remote 
host. The analysis has previously been investigated in the context of many-core 
architectures \cite{Varghese2015}, but we believe virtual GPUs can be a better 
option. 

Aggregate risk analysis is performed on a portfolio of risk held by an insurer or reinsurer and provides actuaries and decision makers with millions of alternate views of catastrophic events, such as earthquakes, that can occur and the order in which they can occur in a year. To obtain millions of alternate views, millions of trials are simulated with each trial comprising a set of possible future earthquake events and the probable loss for each trial is estimated. %The inputs required for the analysis, the algorithm to derive risk metrics from the input and the outputs are considered in the following sections.

\subsection{Input and Output Data}
Three data tables are required for the analysis, which are as follows:

i. \textit{Year Event Table}, which is a database of pre-simulated occurrences of events from a catalogue of stochastic events that is denoted as $YET$. Each record in a $YET$ called a `trial', denoted as $T_i$, represents a possible sequence of event occurrences for any given year. The sequence of events is defined by an ordered set of tuples containing the ID of an event and the time-stamp of its occurrence in that trial $T_i = \{(E_{i, 1}, t_{i, 1}), \dots, (E_{i, k}, t_{i, k})\}$.

The set is ordered by ascending time-stamp values. A typical $YET$ may comprise 
thousands to millions of trials, and each trial may have approximately between 
800 to 1500 `event time-stamp' pairs, based on a global event catalogue covering 
multiple perils. The representation of the $YET$ is shown in 
Equation~\ref{equation1}, where $i = 1, 2, \dots$ and $k = 1, 2, \dots, 1500$.

%$YET	= \{ T_i = \{(E_{i, 1}, t_{i, 1}), \dots, (E_{i, k}, t_{i, k})\} \},$ 
%where $i = 1, 2, \dots$ and $k = 1, 2, \dots, 800-1500$.

\begin{equation}
\label{equation1}
YET	= \{ T_i = \{(E_{i, 1}, t_{i, 1}), \dots, (E_{i, k}, t_{i, k})\} \}
\end{equation}

ii. \textit{Event Loss Tables}, which is a representation of collections of specific events and their corresponding losses with respect to an exposure set denoted as $ELT$. Each record in an $ELT$ is denoted as $EL_{i} = \{E_{i}, l_{i}\}$ and the financial terms associated with the $ELT$ are represented as a tuple $\mathcal{I} = (\mathcal{I}_{1}, \mathcal{I}_{2}, \dots)$. 

A typical aggregate analysis may comprise 10,000 $ELTs$, each containing 10,000-30,000 event losses with exceptions even up to 2,000,000 event losses. The $ELTs$ can be represented as shown in Equation \ref{equation2}, where $i = 1, 2, \dots , 30,000$.
%$
%ELT=\left\{
%	\begin{array}{l c l}
%	EL_{i}			&	=	&	\{E_{i}, l_{i}\},\\
%	\mathcal{I} 		&	=	&	(\mathcal{I}_{1}, \mathcal{I}_{2}, \dots)
%	\end{array}\right\}
%$
%with $i = 1, 2, \dots , 10,000-30,000$.
\begin{equation}
\label{equation2}
ELT=\left\{
	\begin{array}{l c l}
	EL_{i}			&	=	&	\{E_{i}, l_{i}\},\\
	\mathcal{I} 		&	=	&	(\mathcal{I}_{1}, \mathcal{I}_{2}, \dots)
	\end{array}\right\}
\end{equation}

iii. \textit{Portfolio}, which is denoted as $PF$ and contains a group of Programs, $P$ represented as
$
PF = \{P_{1}, P_{2}, \dots, P_{n}\}
$
with $n = 1, 2, \dots, 10$.

Each Program in turn covers a set of Layers, denoted as $L$, cover a collection of $ELTs$ under a set of layer terms. A single layer $L_i$ is composed of two attributes. Firstly, the set of $ELTs$
$
\mathcal{E} = \{ELT_1, ELT_2, \dots, ELT_j\}, 
$
and secondly, the Layer Terms, denoted as 
$
\mathcal{T} = (\mathcal{T}_{1}, \mathcal{T}_{2}, \dots).
$

A typical Layer covers approximately 3 to 30 individual $ELTs$ and is represented as shown in Equation \ref{equation3}, where $j = 1, 2, \dots, 30$.
%$
%L=\left\{
%	\begin{array}{l c l}
%	\mathcal{E}	& =	& \{ELT_1, ELT_2, \dots, ELT_j\}, \\
%	\mathcal{T}	& = & (\mathcal{T}_{OccR}, \mathcal{T}_{OccL}, \ mathcal{T}_{AggR}, \mathcal{T}_{AggL})
%	\end{array}\right\}
%$
%with $j = 1, 2, \dots, 3-30$.
\begin{equation}
\label{equation3}
L=\left\{
	\begin{array}{l c l}
	\mathcal{E}	& =	& \{ELT_1, ELT_2, \dots, ELT_j\}, \\
	\mathcal{T}	& = & (\mathcal{T}_{1}, \mathcal{T}_{2}, \dots)
	\end{array}\right\}
\end{equation}

The output of the analysis is a loss value associated with each trial of the $YET$. A reinsurer can derive important portfolio risk metrics such as the Probable Maximum Loss (PML) \cite{pml1} and the Tail Value-at-Risk (TVaR) \cite{tvar1} which are used for both internal risk management and reporting to regulators and rating agencies. Furthermore, these metrics flow into a final stage of the risk analytics pipeline, namely Enterprise Risk Management, where liability, asset, and other forms of risks are combined and correlated to generate an enterprise wide view of risk.

\subsection{Algorithm and GPU Implementation}

Given the above three inputs, Aggregate Risk Analysis is shown in 
Algorithm~\ref{algorithm1}. The data tables, $YET$, $ELT$ and $PF$, are loaded 
into host (CPU) memory. The analysis is performed for each Layer and for each 
Trial in the $YET$ and a Year Loss Table ($YLT$) is produced. In this paper, we 
assume a Portfolio comprising one Program and one Layer, and therefore the for 
loops of lines 1 and 2 iterate once. If there are $N$ available devices (GPUs), 
then the $YET$ is split to $N$ smaller $YETs$, represented as $YET_{i}$, where 
$i=1,2,\dots,N$.

\begin{algorithm}[t]
\caption{Aggregate Risk Analysis}
\label{algorithm1}
\SetAlgoLined
\DontPrintSemicolon

\SetKwInOut{Input}{Input}
\SetKwInOut{Output}{Output}

\BlankLine

\Input{$YET$, $ELT$, $PF$}
\Output{$YLT$}

\BlankLine

\For{each Program, $P$, in $PF$}{
	\For{each Layer, $L$, in $P$}{
		Split $YET$ to $YET_{i}$, where $i=1,2,\dots,N$\;
		\For{each $i$}{
			\textbf{TransferDataToDevice} ($i$, $YET_{i}$, $ELT$)\;
			\textbf{LaunchDeviceKernel} ($i$)\;
		}
	}
}			
Populate $YLT$ from $YLT_{i}$, where $i=1,2,\dots,N$\;
\KwRet\;
\BlankLine
\end{algorithm}

\begin{algorithm}[b]
\caption{TransferDataToDevice Function}
\label{algorithm2}
\SetAlgoLined
\DontPrintSemicolon

\SetKwInOut{Input}{Input}
\SetKwInOut{Output}{Output}

\BlankLine

\Input{$i$}
\BlankLine
Select device $i$\;
Copy $YET_{i}$, $ELT$ to device $i$\;
\KwRet\;
\end{algorithm}

\begin{algorithm}[t]
\caption{LaunchDeviceKernel Function}
\label{algorithm3}
\SetAlgoLined
\DontPrintSemicolon

\SetKwInOut{Input}{Input}
\SetKwInOut{Output}{Output}

\BlankLine

\Input{$i$}
\Output{$YLT_{i}$}
\BlankLine

Select device $i$\;
\For{each Trial, $T$, in $YET_{i}$}{
	\For{each Event, $E$, in $T$}{
		\For{each $ELT$ covered by $L$}{
			Lookup $E$ in the $ELT$ and find corresponding loss, $l_{E}$\;
			Apply Financial Terms to $l_{E}$\;
			$l_{T} \leftarrow$ $l_{T}$ + $l_{E}$\;
		}
		Apply Financial Terms to $l_{T}$\;	
	}
}
\KwRet\;
\end{algorithm} 

There are two functions that facilitate device execution. The first function \texttt{TransferDataToDevice} copies $YET_{i}$ and the $ELT$ to the device memory as shown in Algorithm~\ref{algorithm2}.

The second function \texttt{LaunchDeviceKernel} executes the function on the device as shown in Algorithm~\ref{algorithm3}. Each event of a trial and its corresponding event loss in the set of $ELTs$ associated with the Layer is determined. A set of contractual financial terms ($\mathcal{I}$) are applied to each loss value of the Event-Loss pair extracted from an $ELT$ to the benefit of the layer. The event loss for each event occurrence in the trial, combined across all $ELTs$ associated with the layer, are subject to further financial terms ($\mathcal{T}$) \cite{Varghese2012}. 

Two occurrence terms, namely (i) Occurrence Retention, $\mathcal{T}_{OccR}$, which is the retention or deductible of the insured for an individual occurrence loss, and (ii) Occurrence Limit, $\mathcal{T}_{OccL}$, which is the limit of coverage the insurer will pay for occurrence losses in excess of the retention are applied. Occurrence terms are applicable to individual event occurrences independent of any other occurrences in the trial. The event losses net of occurrence terms are then accumulated into a single aggregate loss for the given trial. The occurrence terms are applied as $l_{T} = min ( max ( l_{T} - \mathcal{T}_{OccR} ), \mathcal{T}_{OccL})$.

Two aggregate terms, namely (i) Aggregate Retention, $\mathcal{T}_{AggR}$, which is the retention or deductible of the insured for an annual cumulative loss, and (ii) Aggregate Limit, $\mathcal{T}_{AggL}$, which is the limit or coverage the insurer will pay for annual cumulative losses in excess of the aggregate retention are applied. Aggregate terms are applied to the trial's aggregate loss for a layer. The aggregate loss net of the aggregate terms is referred to as the trial loss or the year loss. The aggregate terms are applied as $l_{T} = min ( max ( l_{T} - \mathcal{T}_{AggR} ), \mathcal{T}_{AggL})$.

A single thread is employed for the computations of each trial of the application. $ELTs$ corresponding to a Layer were implemented as direct access tables to facilitate fast lookup of losses corresponding to events. Each $ELT$ is implemented as an independent table; therefore, in a read cycle, each thread independently looks up its events from the $ELTs$. All threads within a block access the same $ELT$. The device global memory stores all data required for the analysis. Chunking, which refers to processing a block of events of fixed size (or chunk size) for the efficient use of shared memory is employed to optimise the implementation; the computations related to the events in a trial and for applying financial terms benefit from chunking. The financial terms are stored in the streaming multi-processor's constant memory. In this case, chunking reduces the number of global memory update and global read operations.

In this paper, the implementation of fine-grain parallelism in \texttt{LaunchDeviceKernel} is not the focus. Instead, the optimisation of performance and efficiency of resource utilisation by managing the two functions, namely \texttt{TransferDataToDevice} and \texttt{LaunchDeviceKernel} on virtual GPUs is considered and reported in the next section.

\section{Evaluation}
\label{evaluation}
In this section we optimise the performance of the financial risk application to reduce its execution time such that real-time response can be achieved. To this end we present (i) the hardware platform on which the experiments are performed and, (ii) the use of the remote GPU virtualisation framework, and (iii) an approach for transferring data from a CPU to GPUs with the aim of reducing the execution time.

\subsection{Platform}
\label{platform}
The experimental platform employed in this research comprises 1027GR-TRF Supermicro nodes. Each node contains two Intel Xeon E5-2620 v2 processors, each with six cores, operating at 2.1 GHz and 32 GB of DDR3 SDRAM memory at 1600 MHz. Each node has a Mellanox ConnectX-3 VPI single-port InfiniBand adapter (InfiniBand FDR) as well as a Mellanox ConnectX-2 VPI single-port adapter (InfiniBand QDR). The nodes are connected either by a Mellanox switch MTS3600 with QDR compatibility (a maximum rate of 40Gb/s) or by a Mellanox Switch SX6025, which is compatible with InfiniBand FDR (a maximum rate of 56Gb/s). One NVIDIA Tesla K20 GPU is available for acceleration on each node. Additionally, one SYS7047GR-TRF Supermicro server with identical processors was populated with 4 NVIDIA Tesla K20 GPUs and 128 GB of DDR3 SDRAM memory at 1600MHz, to serve as a local server for the purpose of comparison. The CentOS 6.4 operating system was used, and the Mellanox OFED 2.4-1.0.4 (InfiniBand drivers and administrative tools) was used at the servers along with CUDA 6.5.

\subsection{Application Scalability}
\label{cuda_scalability}

As presented in Section \ref{introduction} the use of multiple GPUs reduces 
the execution time of the application by evenly distributing computations across the GPUs assigned to the application. However, a closer look at the 
performance as shown in Figure~\ref{cuda_performance} highlights that the scalability of the application as the number of GPUs increases is sub-linear. Table~\ref{scalability_application} is the result of executing the application on the Supermicro SYS7047GR-TRF server using CUDA with up to four GPUs. The normalised execution time indicates that perfect scalability is not achieved. For example, when two GPUs are used the normalised execution time should be 0.5 instead of 0.506 and similarly when four GPUs are employed 0.25 is expected as against 0.261. The offset of execution time with respect to perfect scalability as a reference increases with the number of GPUs involved in the computations.

\begin{table*}[t]
\caption{Scalability of the financial risk application when executed using CUDA}
%\vspace{-0.25cm}
\label{scalability_application}
\begin{center}
\begin{tabular}{ | l | c | c | c |}
\hline
  &  \multicolumn{3}{|c|}{No. of GPUs}  \\ \cline{2-4}
  &  1 GPU &   2 GPUs & 4 GPUs \\ \hline \hline
Total execution time & 10.928 & 5.53 & 2.857 \\ \hline
Normalised execution time & 1 & 0.506 &  0.261 \\ \hline
Execution time with perfect scalability & 10.928 & 5.464 &  2.732 \\ \hline
Offset with respect to perfect scalability & 0 & 0.066 & 0.125 \\ \hline
\% offset with respect to perfect scalability & 0 & 1.2\% & 4.57\% \\ \hline
\end{tabular}
\end{center}
\end{table*}

\begin{figure}[t]
\centering
\includegraphics[width=0.5\textwidth]{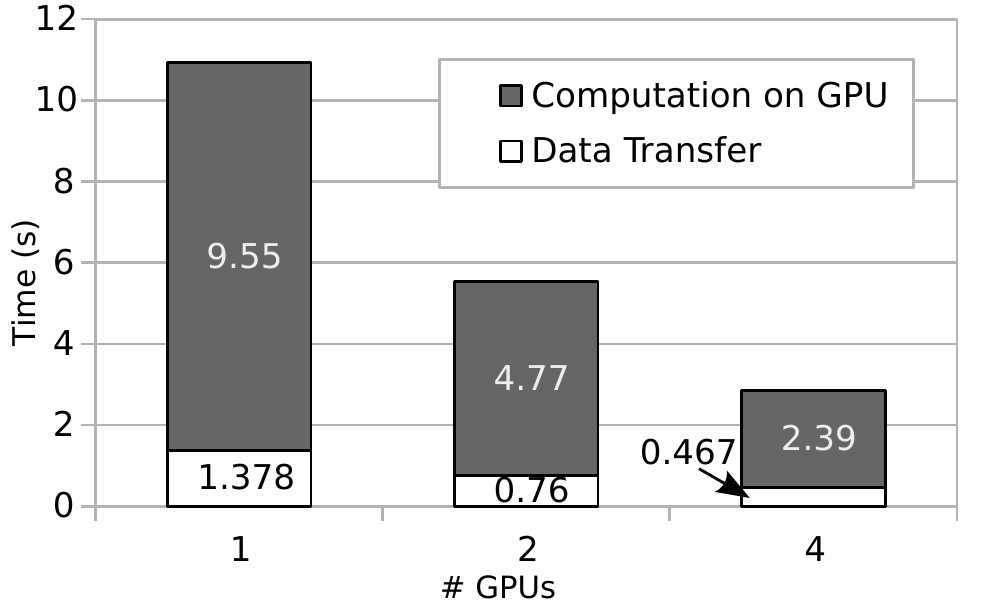}
\caption{Computation and data transfer times for the financial 
risk application when executed on single and multiple GPUs with CUDA}
\label{CUDA_execution_times_separated}
\end{figure}

To account for sub-linear scalability further investigations were carried out. The time taken for computations on the GPUs and the time 
taken for transferring data to the GPUs (1, 2, and 4 GPUs) were considered as shown in Figure~\ref{CUDA_execution_times_separated}. The GPU computations take most of the execution time of the application (87.39\%, 86.25\%, and 63.65\% of the total application execution time when 1, 2, and 4 GPUs are used respectively). The GPU computations scale in a perfect manner as the number of GPUs available to the application is increased. However, the time taken for data transfer does not scale well and accounts for 12.6\%, 13.74\%, and 16.34\% of total execution time when 1, 2, and 4 GPUs are used, respectively. 

At first glance, it can be assumed that the increase in data transfer time may be due to the lower communication bandwidth of CUDA for transfers of small chunks of data (refer Figure~\ref{figure5c} and Figure~\ref{figure5d}). When pageable memory is transferred the attained bandwidth for data smaller than 10~MB is significantly reduced. Therefore, given that the size of input data transferred to each GPU is 
progressively reduced as the number of GPUs increases, then the input data may be smaller than 10~MB and thus the effective bandwidth for moving data to the GPUs is reduced in practice. However, in the case of our application the initial data size is 4~GB and when this data is shared among four GPUs the data transferred to each GPU is larger than 10~MB. Hence, the data transfer to the GPUs is performed at full bandwidth. 

\begin{figure*}[t!]
\centering
	\subfloat[Data transferred to each GPU]
	{	\label{transfer_per_GPU}
		\includegraphics[width=0.48\textwidth]
		{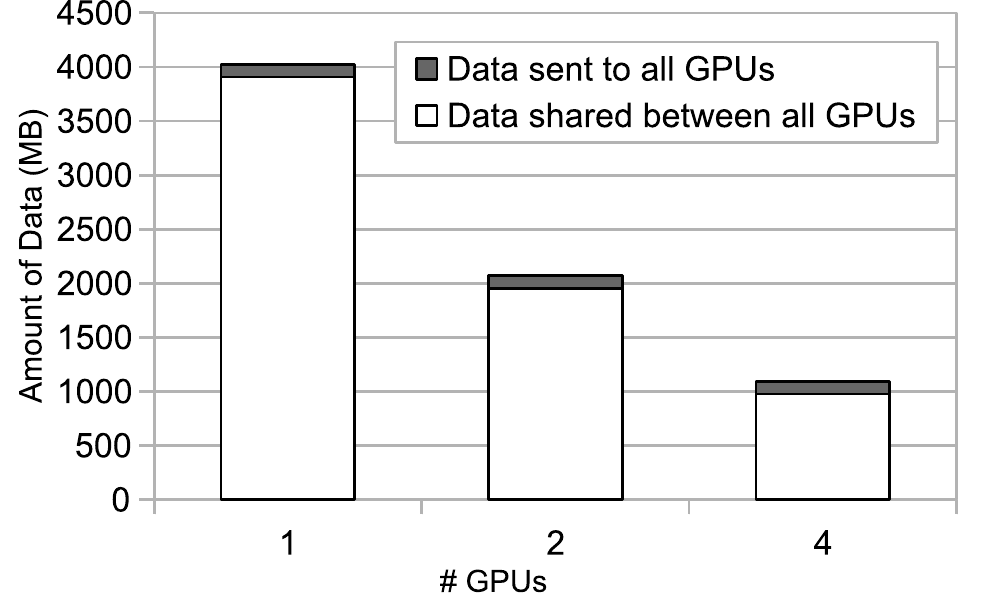}
	}
	\hfill
	\subfloat[Total data transferred to all GPUs]
	{	\label{total_transfers}
		\includegraphics[width=0.48\textwidth]
		{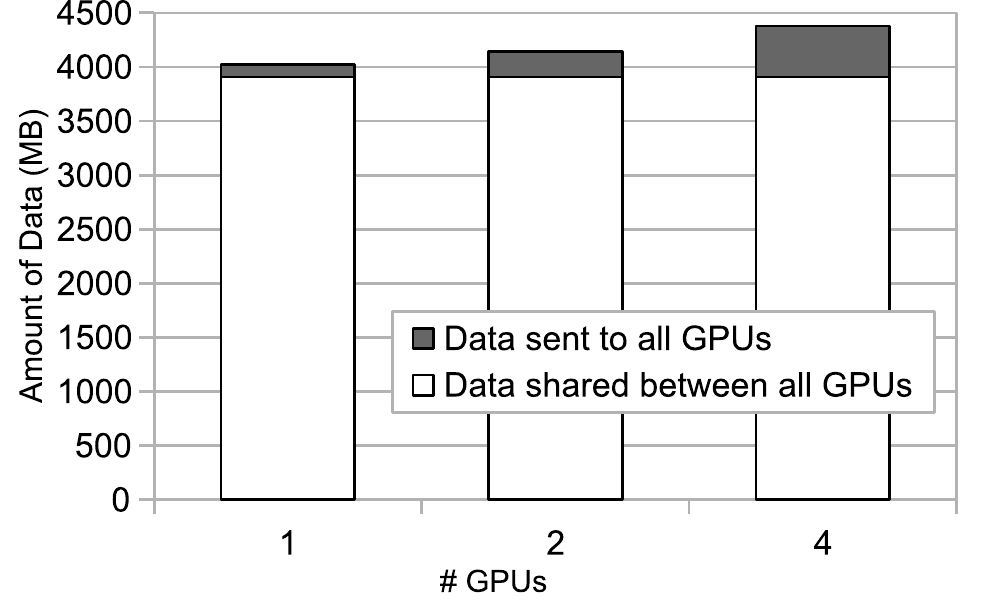} 
	} \\
	\caption{Amount of data transferred during the execution of the financial risk application}
	\label{amount_of_transferred_data}
\end{figure*}

A closer look at the application reveals that the $YET$ data structure (4~GB) presented in Section~\ref{application} is uniformly split between the GPUs for computations. However, the $ELTs$ and $PF$ data structures (120~MB and 4~MB) are not split between the GPUs, instead are transferred fully to each GPU. Consequently, the total data movement to GPUs increases which is shown in Figure~\ref{amount_of_transferred_data}. Excluding the $ELTs$, the data that is not split between the GPUs is less than 10~MB resulting in a lower bandwidth for transferring this data requiring an additional 2.6 milliseconds. However, this cannot fully account for sub-linear performance.  

\begin{figure}[t]
\centering
	\includegraphics[width=0.5\textwidth]{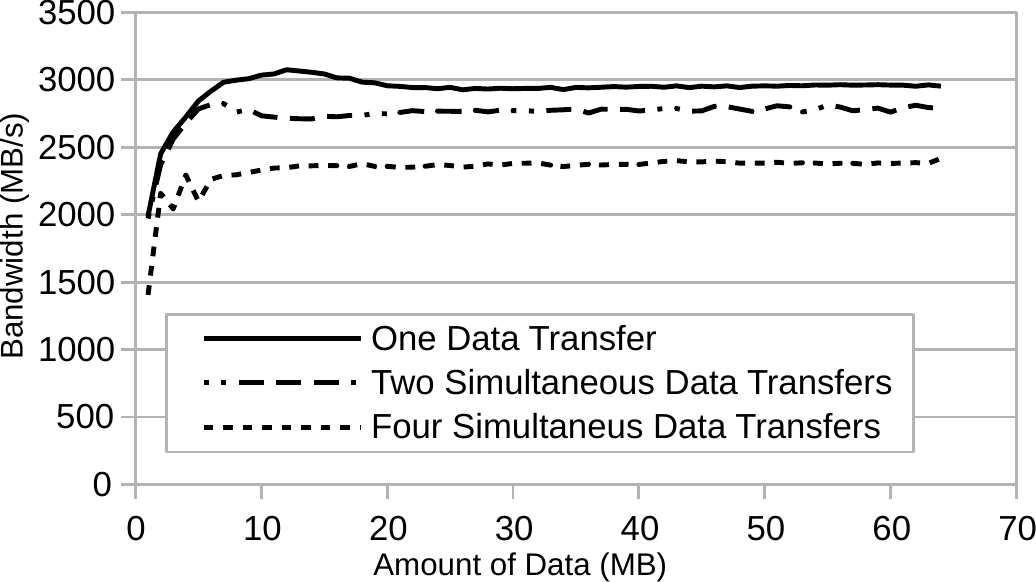}
	\caption{Attained bandwidth when concurrent data transfers 
to GPUs are performed. Source data is located in the same memory bank.}
	\label{memory_bottleneck}
\end{figure}

One important reason for the degradation of performance is data transfers to all GPUs are concurrently performed. Although each GPU is located in a different PCIe link, all data is extracted from main memory, which results in a bottleneck. This memory bottleneck is highlighted in Figure~\ref{memory_bottleneck}, which shows the bandwidth attained for each individual data copy when several data transfers are carried out concurrently to different destination GPUs by a single memory controller. 

We summarise that for the financial risk application executing on multiple GPUs data transfers do not scale perfectly as the computations for two reasons. Firstly, there are input data structures that cannot be split between the GPUs and need to be copied onto each GPU creating an overhead. Secondly, concurrent data transfers from the CPU main memory to GPUs result in a bottleneck at the memory controller. 

\subsection{Reducing Execution Time Using rCUDA}
\label{rcuda_scalability}

\begin{figure*}[t!]
\centering
\subfloat[On QDR InfiniBand]
{\label{scalability_rCUDA-QDR}
\includegraphics[width=0.48\textwidth]
{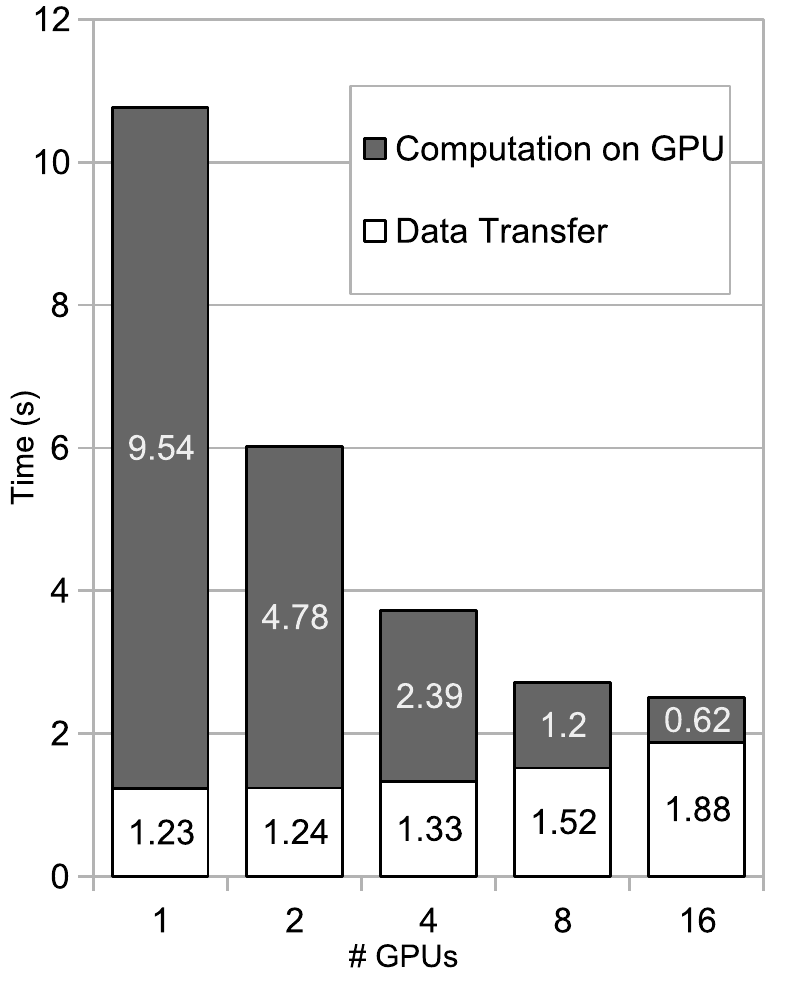}
}
\hfill
\subfloat[On FDR InfiniBand]
{\label{scalability_rCUDA-FDR}
\includegraphics[width=0.48\textwidth]
{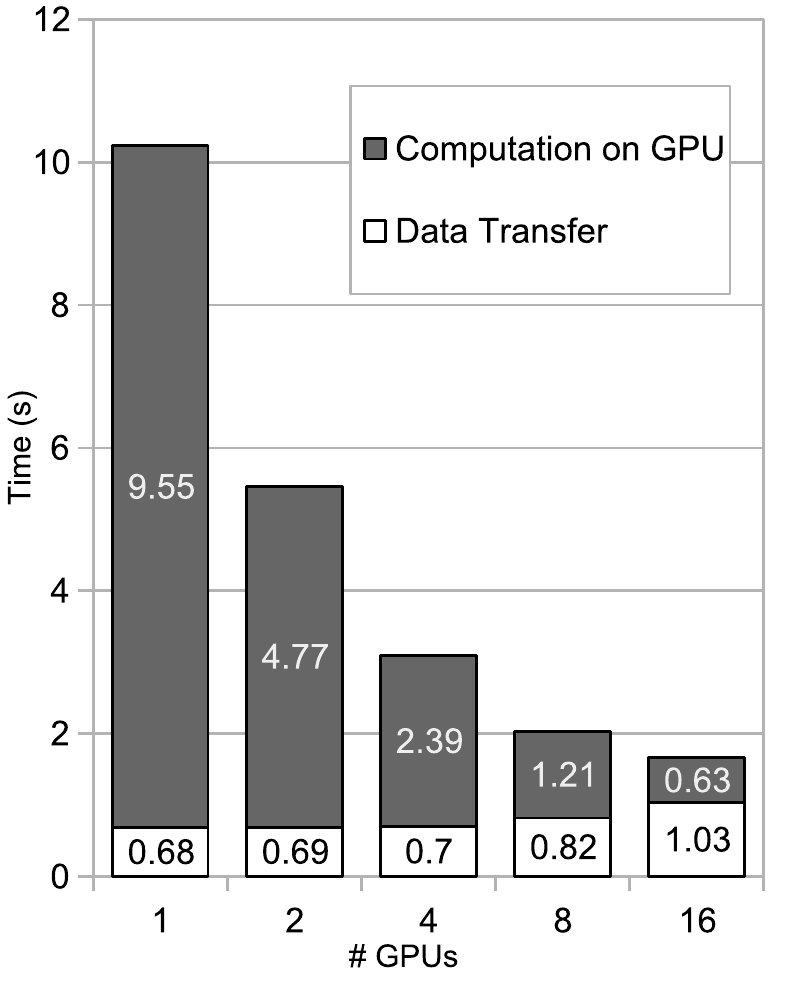} 
} \\
\caption{Scalability of the financial risk application when executed with rCUDA.}
\label{scalability_with_rCUDA}
\end{figure*}

Current servers are constrained in the number of GPUs that can be accommodated on them\footnote{Manufacturers, such as Cirrascale and Supermicro, have integrated up to 8 GPU cards in a single server. However, these are exceptions and costly options. Moreover, there are performance bottlenecks since the GPUs are usually grouped as a set of four cards that share a single PCIe x16 link with a processor socket. This results in slower communication between main memory and the GPUs. Performance is further degraded when a GPU card comprises multiple devices.}. We believe remote GPU virtualisation (in this research rCUDA is employed) is an appropriate mechanism to make a large number of GPUs available to an application. 
Figure~\ref{scalability_rCUDA-QDR} and  Figure~\ref{scalability_rCUDA-FDR} present the performance of the application using the QDR InfiniBand and the FDR InfiniBand networks respectively for up to 16 GPUs. 

Figure~\ref{scalability_with_rCUDA} indicates that the computation times when using rCUDA on 1, 2, and 4 GPUs are the same as shown in 
Figure~\ref{CUDA_execution_times_separated} using CUDA. This is expected given that the computation time on the GPU is independent of whether it is on the same node as the application or on a remote node. With increasing number of GPUs there is perfect scalability. When 16 GPUs are employed, the computation time is less than one second (0.62 seconds) making it possible to do an industry size simulation in real-time. 

Two observations are made regarding data transfers. Firstly, when one remote GPU is used, the data transfer time using rCUDA is better than using CUDA (CUDA requires 1.378 seconds whereas rCUDA takes 1.23 seconds with QDR InfiniBand and 0.68 seconds with FDR InfiniBand). This lower transfer time as considered in Figure~\ref{figure5c} is because rCUDA obtains more bandwidth than CUDA by using pageable memory. The improvement of communication performance is seen in Figure~\ref{scalability_rCUDA-FDR} for 2 GPUs. 

\begin{figure*}[t!]
\centering
	\subfloat[On QDR InfiniBand]
	{	\label{NIC_bottleneck-QDR}
		\includegraphics[width=0.48\textwidth]
		{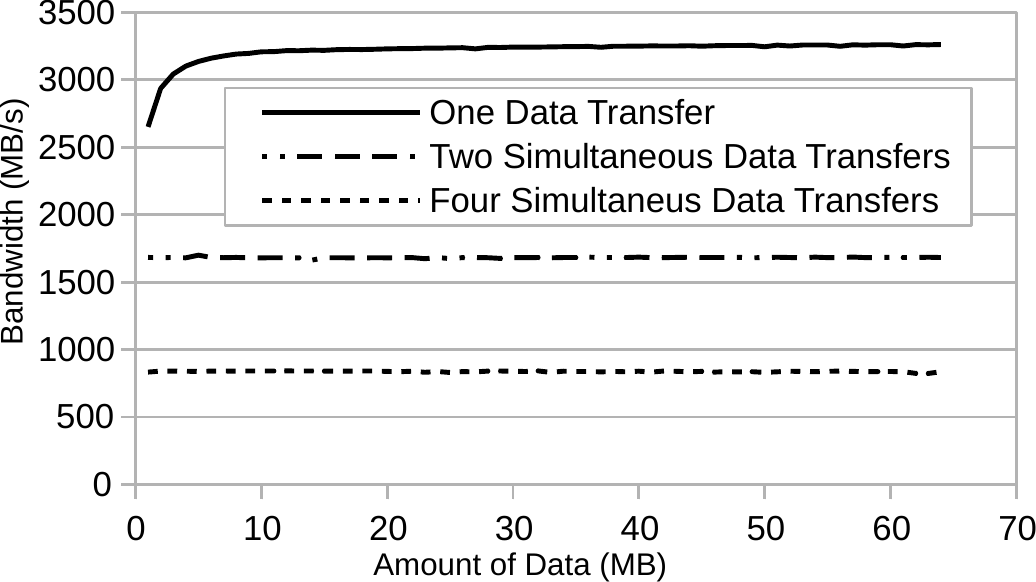}
	}
	\hfill	
	\subfloat[On FDR InfiniBand]
	{	\label{NIC_bottleneck-FDR}
		\includegraphics[width=0.48\textwidth]
		{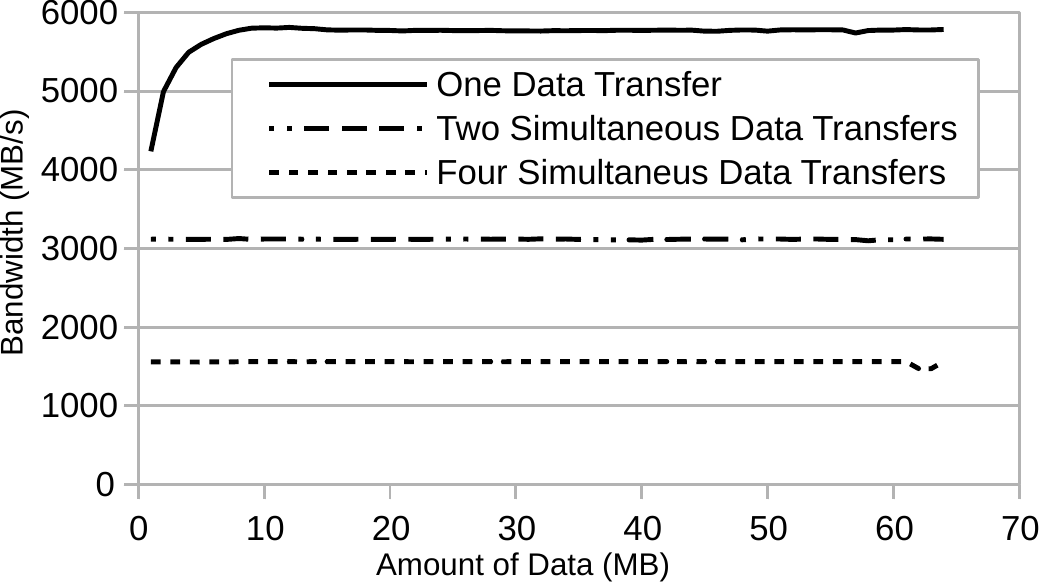} 
	} \\
	\caption{Bandwidth attained for multiple data transfers concurrently to different remote GPUs using rCUDA.}
\label{NIC_bottleneck}
\end{figure*}

Secondly, data transfer using rCUDA follows a different trend to CUDA. For CUDA the data transfer times to each GPU reduced as the number of GPUs increased (refer Figure~\ref{CUDA_execution_times_separated}). On the contrary, rCUDA time increases when both QDR and FDR InfiniBand are used. This is not surprising since the reasons for sub-linear scalability of data transfer time considered in the previous section is applicable for rCUDA. In this case, the bandwidth bottleneck is the InfiniBand card in the cluster node executing the application, which is a single communication link for all the GPUs. This bottleneck is highlighted in Figure~\ref{NIC_bottleneck}. 

Figure~\ref{NIC_bottleneck} shows the bandwidth achieved for individual data transfer to a different remote GPU when multiple transfers are executed concurrently. The bandwidth for each transfer is proportional to the number of data movement operations in progress. In addition to the previous observations that result in an increase of data transfer times, there are a large number of {\tt cudaMalloc()} functions that are invoked prior to the data transfer (the memory allocation time is included in the data transfer time). In rCUDA, memory allocations for a large number of data structures on remote GPUs requires 2.7~milliseconds with FDR InfiniBand (compared to 1.7~milliseconds in CUDA on a local GPU) and 2.67~milliseconds with QDR InfiniBand (lower time due to low latency, despite reduced bandwidth \cite{Reano2013}). Therefore, when a large number of GPUs are used by an application the time required for memory allocations can increase up to 43.2~milliseconds for 16 remote GPUs; this is 4.2\% of the total data transfer time.

The use of rCUDA allows to leverage a large number of GPUs to speed up the application despite poor performance for data transfers. The total execution time is reduced from 2.86 seconds when using local GPUs on CUDA to 1.66 seconds when using remote GPUs on rCUDA. Reducing the total execution time enables the application to provide a solution in real-time. 

\subsection{Mitigating the Impact of Data Transfers in rCUDA}

In this section, we consider two data transfer modes, namely concurrent and sequential, and further develop an approach based on multi-tenant GPUs in rCUDA.  

\subsubsection{Concurrent vs Sequential Data Transfers}

\begin{figure*}[t!]
	\centering
	\subfloat[Concurrent data transfers]
	{	\label{concurrent_communications}
		\includegraphics[width=1.0\textwidth]
		{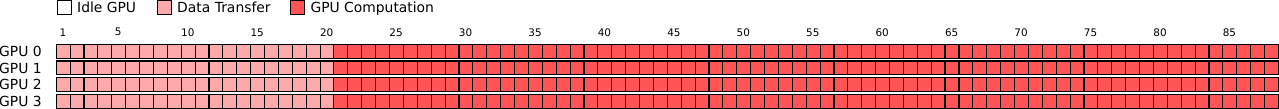}
	}

	\subfloat[Sequential data transfers]
	{	\label{sequential_communications_1vgpu}
		\includegraphics[width=1.0\textwidth]
		{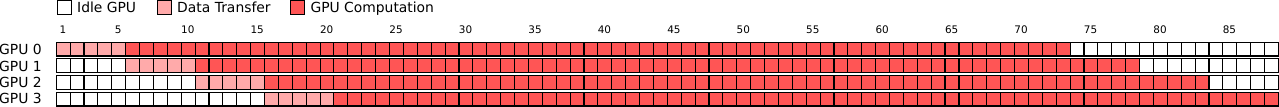} 
	} \\
	\caption{Communication approaches for transferring data to GPUs.}
	\label{communication_models}
\end{figure*}

Figure~\ref{concurrent_communications} shows the life cycle of execution of a real application using rCUDA with four remote GPUs and FDR InfiniBand. Each cell represents execution time of 35~milliseconds. This corresponds to the four GPU execution shown in Figure~\ref{scalability_rCUDA-FDR}. The same amount of data is moved to the four GPUs concurrently by interleaving across the network and the remote GPUs start computations at the same time approximately. However, from Figure~\ref{NIC_bottleneck} it was noted that the bandwidth achieved is inversely proportional to the number of multiple data transfers concurrently performed which results in degrading performance.

An alternate method is shown in Figure~\ref{sequential_communications_1vgpu}. Data to the first GPU is transferred without sharing the bandwidth for the remaining three data streams. Since there is no competition for bandwidth it only takes a quarter of the time required when data is concurrently transferred (shown in Figure~\ref{concurrent_communications}). Computations on the first GPU start while data is transferred to the second GPU. In this manner, data transfer is performed on fully available network bandwidth. This is referred to as the sequential data transfer method.  

\begin{figure}[t!]
\centering
\subfloat[Concurrent]
{\label{energy-utilization-conc}
\includegraphics[width=0.5\textwidth]
{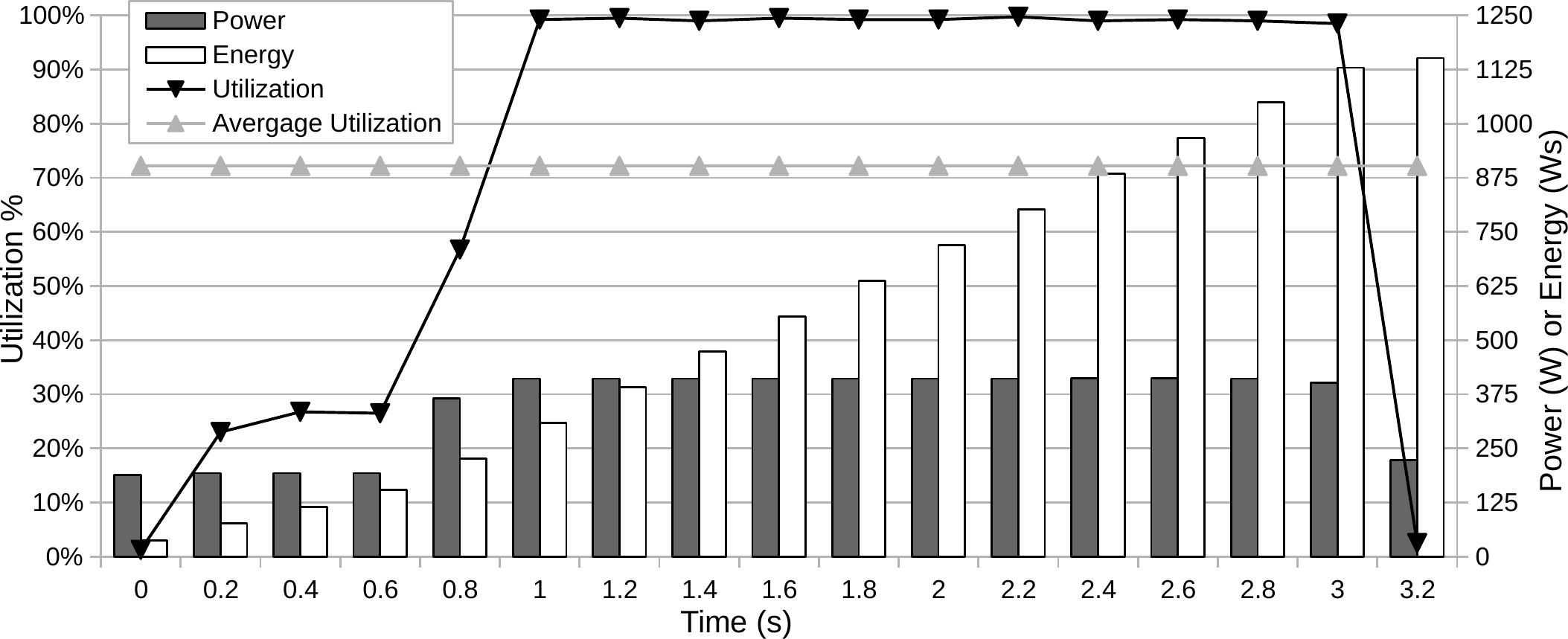}
}

\subfloat[Sequential]
{\label{energy-utilization-seq1vgpu}
\includegraphics[width=0.5\textwidth]
{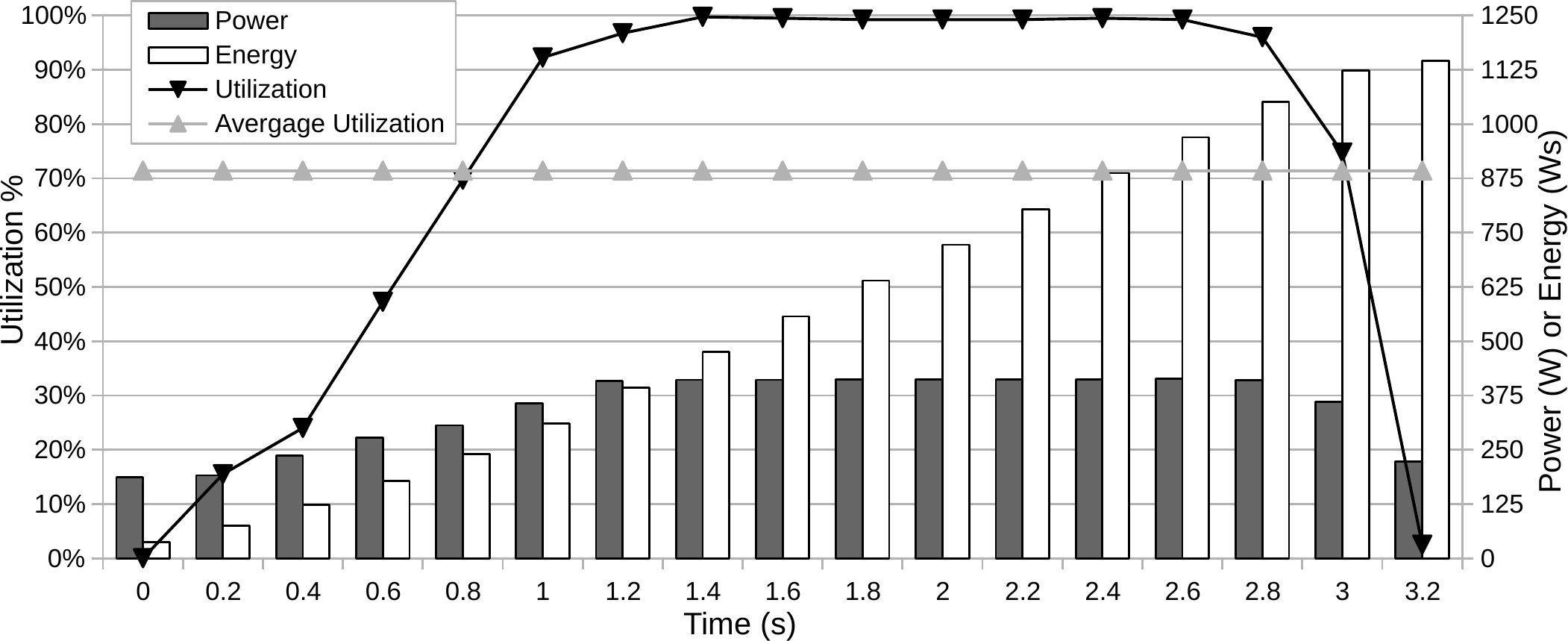} 
} \\
\caption{GPU utilisation, power and energy consumption of concurrent and sequential data transfers to GPUs considered in Figure~\ref{communication_models}}
\label{energy-utilization-conc-seq1vgpu}
\end{figure}

Data is transferred at full network bandwidth and there is an overlap with GPU computations in the sequential data transfer approach. However, it is noted that the execution time is not reduced since the fourth GPU begins its computations when it would in concurrent data transfers. Figure~\ref{energy-utilization-conc-seq1vgpu} shows the GPU utilisation, power and energy consumption of concurrent and sequential data transfers to GPUs. The average values of the four GPUs considered in Figure~\ref{communication_models} are used. The Y-axis on the left indicates GPU utilisation and the Y-axis on the right shows power (in Watts) and energy (in Watts per second, denoted as Ws in the figure) consumed. The power and energy of GPUs are measured instead of the cluster since multiple GPU configurations ($n$ GPUs per node) could be employed, which results in different energy measurements. There are no gains in the energy consumed and very little difference in GPU utilisation for both concurrent and sequential transfers. 

Regardless, in this research sequential data transfer is foundational in developing an optimised approach for executing the application using remote GPUs which is based on multi-tenancy of virtual GPUs. 

\begin{figure*}[t!]
\centering
	\subfloat[2 vGPUs per GPU]
	{	\label{sequential_communications_2vgpu}
		\includegraphics[width=1.0\textwidth]
		{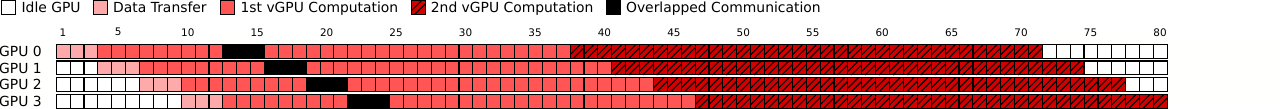}
	}

	\subfloat[4 vGPUs per GPU]
	{	\label{sequential_communications_4vgpu}
		\includegraphics[width=1.0\textwidth]
		{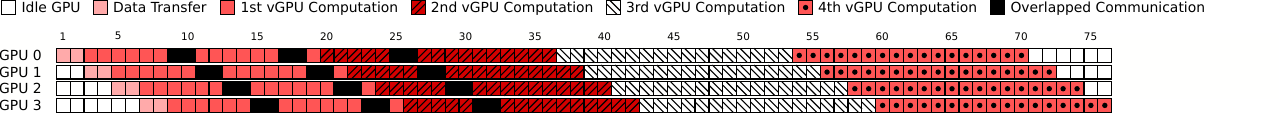} 
	} \\
	\caption{Sequential data copies with several vGPUs per GPU.}
	\label{sequential_communications}
\end{figure*}

\subsubsection{Multi-tenancy Approach}

The key concept of the multi-tenancy approach is based on the fact that current GPUs perform kernel executions and DMA (Direct Memory Access) operations concurrently. If it were possible to move data to a GPU the same time it was executing a kernel, there could be gains in further improving the performance of the executing application. 

This can be facilitated by a multi-tenancy approach in which a number of remote GPUs (or virtual GPUs referred to as vGPUs) reside on or are mapped onto the same physical GPU (pGPU)\footnote{Multi-tenancy is achieved on rCUDA by setting two environment variables prior to application execution, namely {\small  \tt RCUDA\_DEVICE\_COUNT} and {\small \tt RCUDA\_DEVICE\_j}. The first variable indicates the number of GPUs accessible to the application. The second variable indicates the cluster node in which the $j^{th}$ GPU is located. For example, ``{\small \tt export RCUDA\_DEVICE\_COUNT=2}'' when $2$ GPUs are assigned to the application and ``{\small \tt export RCUDA\_DEVICE\_0=192.168.0.1}'' and ``{\small \tt export RCUDA\_DEVICE\_1=192.168.0.2}''. 
The server of the {\small \tt RCUDA\_DEVICE\_j} variables need to point to the same node. Hence, the application does not require to be modified to accommodate multi-tenancy using rCUDA.}. Figure~\ref{sequential_communications} shows the concept of multi-tenancy when 2 and 4 vGPUs are mapped to a pGPU.

When 2 vGPUs are mapped on to a pGPU as shown in Figure~\ref{sequential_communications_2vgpu} 8 GPUs are available to the application (4 pGPUs are used). Input data will be split such that 8 GPUs will be used for computations. The initial data transfer is shown as {\em``Data Transfer''} followed by computations by the first vGPU labelled as {\em ``1st vGPU Computation´´}. After transferring data in the 12th time step, there are four more vGPUs that will require their input data. Data transferred to the remaining four vGPUs beginning at time step 13 are overlapped with the computations of the first four vGPUs. Since two vGPUs are mapped onto a single pGPU, computations of both vGPUs cannot progress in parallel as they belong to different GPU contexts. Therefore, the NVIDIA driver executes them sequentially (using as many GPU resources required by each kernel). So the second kernel must wait until the execution of the first kernel is completed.

Two key observations are made from multi-tenant executions. Firstly, the total execution time has reduced in contrast to the execution life cycle presented in Figure~\ref{sequential_communications_1vgpu} although the same hardware resources are used. The application completed execution in time step 80 using 2 vGPUs per pGPU compared to time step 88 when no multi-tenancy is employed. The time that each GPU computes is exactly the same. The time saved is because of the overlap between computations and data transfers of multiple vGPUs on the same pGPU. In Figure~\ref{sequential_communications_1vgpu} data transfers overlapped with computations of other pGPUs but there were no overlaps on the same GPU. 

Secondly, the data transfer time takes longer when more vGPUs are employed. In Figure~\ref{sequential_communications_1vgpu}, 
data is transferred completely to all GPUs at time step 20, whereas in Figure~\ref{sequential_communications_2vgpu}, the input data arrives at time step 24. The reasons for longer data transfer times have been considered in the previous section. Despite the larger data transfer time, the total execution time gains since there is an overlap between computation and data movement.

\begin{figure}[t!]
\centering
\subfloat[2 vGPUs per pGPU]
{\label{energy-multi-tenancy-2vgpu}
\includegraphics[width=0.5\textwidth]
{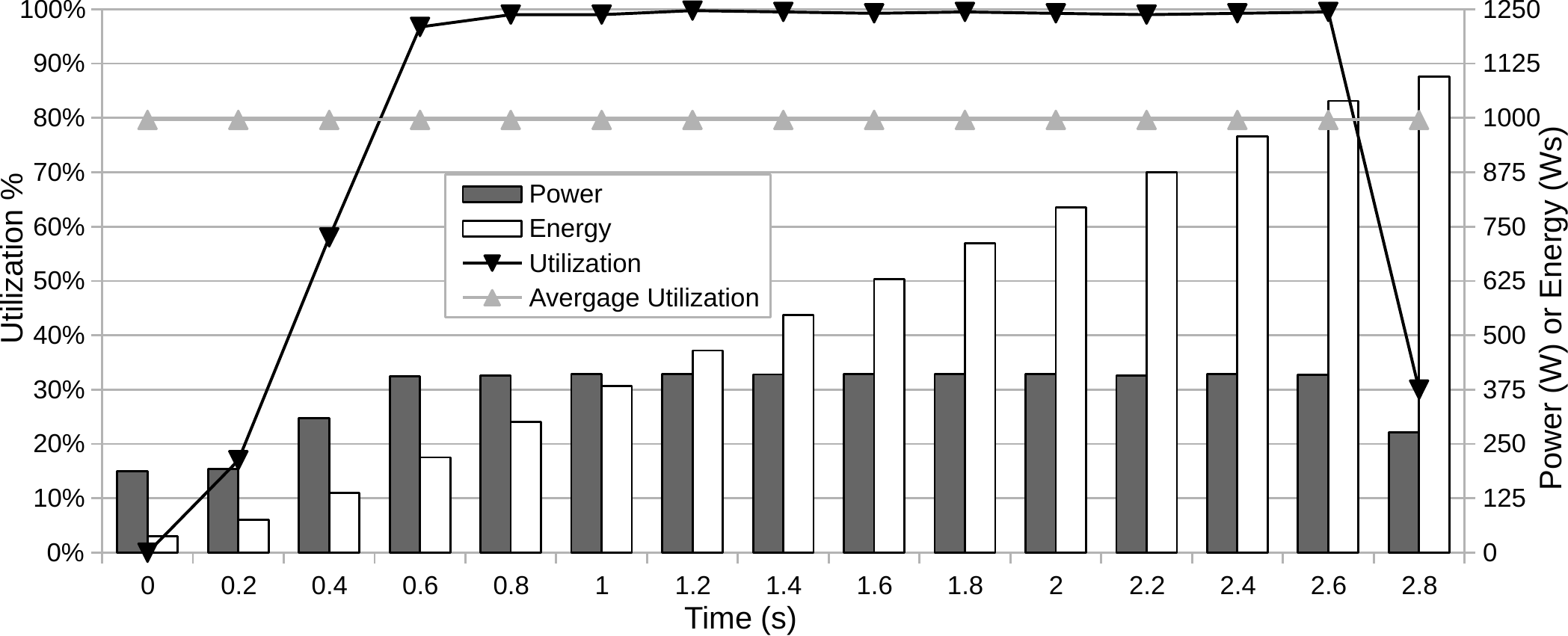}
}

\subfloat[4 vGPUs per pGPU]
{\label{energy-multi-tenancy-4vgpu}
\includegraphics[width=0.5\textwidth]
{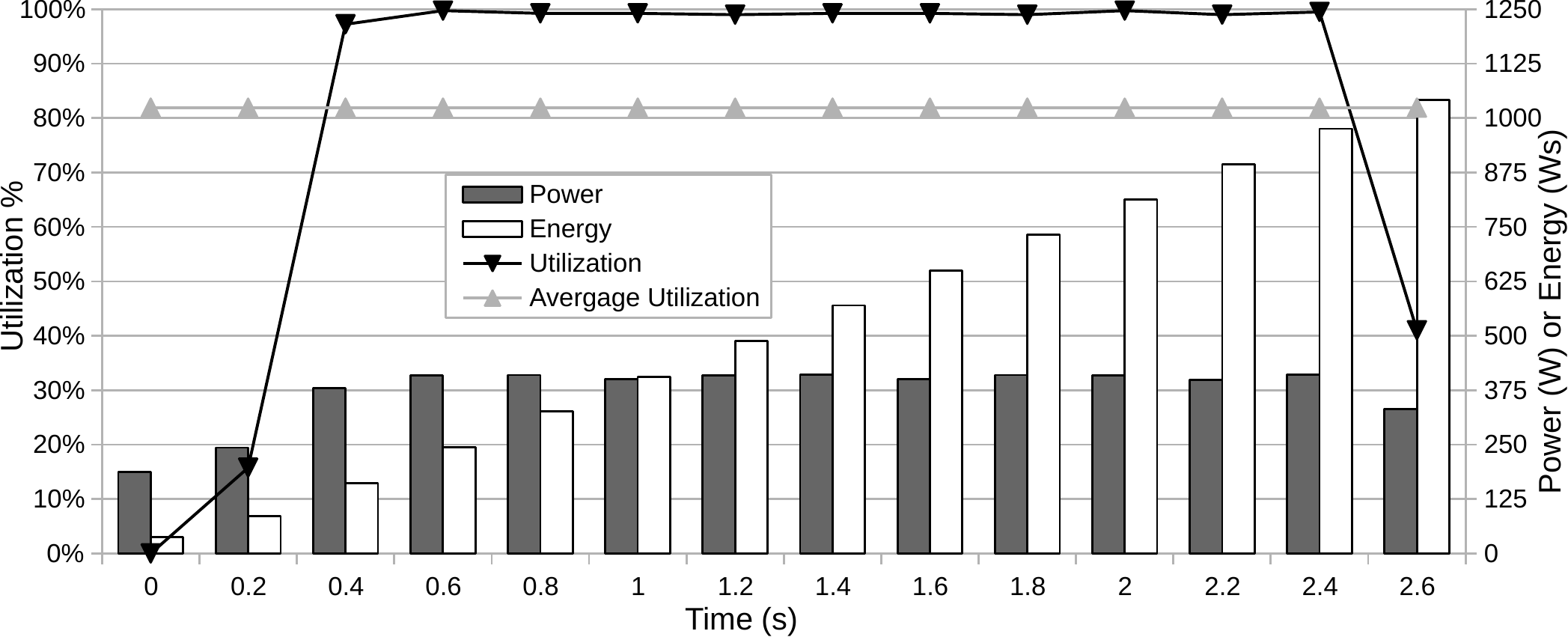} 
} \\
\caption{GPU utilisation, power and energy consumption of the multi-tenancy approach considered in Figure~\ref{sequential_communications}.}
\label{energy-multi-tenancy}
\end{figure}

Figure~\ref{sequential_communications_4vgpu} shows the use of 16 vGPUs mapped on to 4 pGPUs. The execution time is further reduced due to the larger overlap between computation and data transfers when compared to 2 vGPUs residing on a single pGPU. Again the time for computing is the same on each physical GPU but the data copying time has increased. The overall execution time is further reduced to 76 time steps. 

Multi-tenancy can be analysed from the perspective of energy required to complete the execution of the application. Figure~\ref{energy-multi-tenancy} shows the energy consumed during the execution of the application along with the utilization of the physical GPU. The multi-tenancy energy consumption is lower than sequential communications without an overlap between data transfers and computations on the same GPU seen in Figure~\ref{energy-utilization-conc-seq1vgpu}. The energy consumed is 1145 Watts per second without using multi-tenancy and 1094 and 1041 Watts per second when 2 and 4 vGPUs are tenants on a pGPU, respectively. It is observed that GPU utilisation increases in the multi-tenancy approach. The average GPU utilisation rises from 71.44\% without multi-tenancy up to 79.65\% for 2 vGPUs per pGPU and up to 81.93\% when 4 vGPUs are mapped on to a pGPU.

In short, multi-tenancy allows for data transfers to be overlapped with computations on the same GPUs thereby reducing total execution time of the financial risk application. Furthermore, the energy required to execute the application is reduced and the GPU utilisation is increased.

\subsection{Performance Analysis Using Multi-tenancy}

\begin{figure*}[t!]
	\centering
		\includegraphics[width=1.0\textwidth]{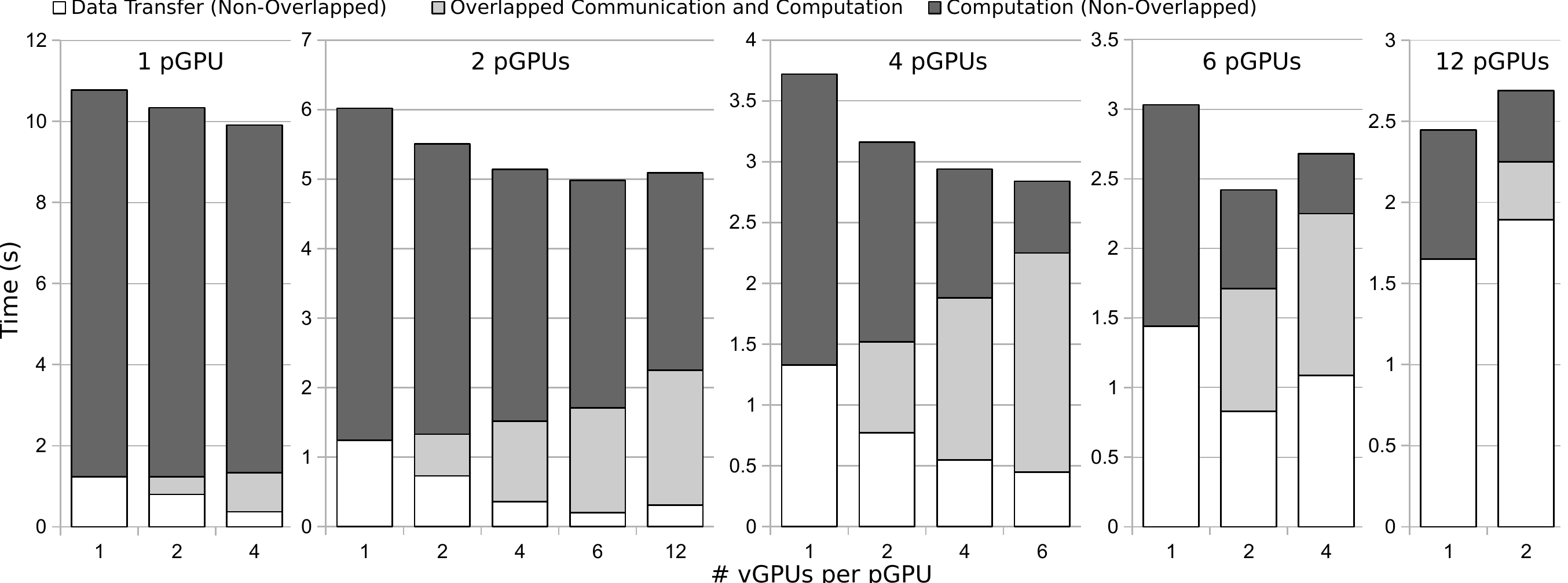}
		\caption{Application performance for different combinations of pGPUs and vGPUs using QDR InfiniBand}
	\label{perf_analysis_QDR}
\end{figure*}

\begin{figure*}[t!]
	\centering
		\includegraphics[width=1.0\textwidth]{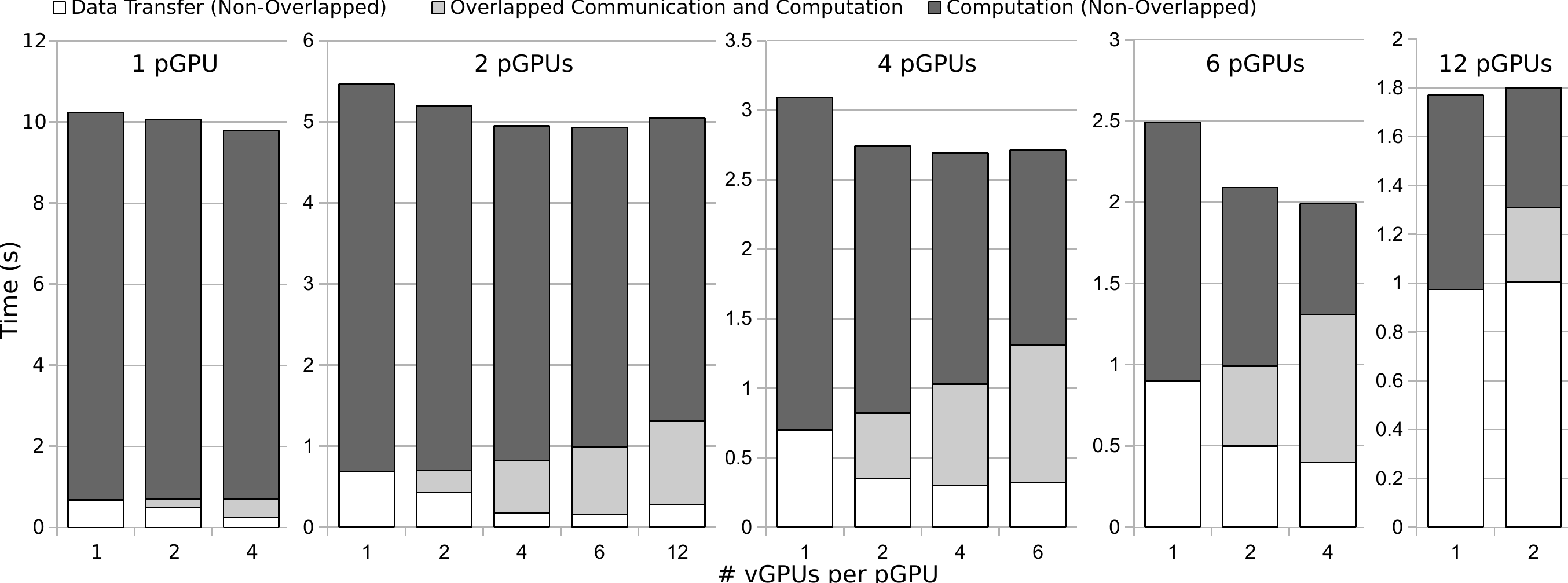}
		\caption{Application performance for different combinations of pGPUs and vGPUs using FDR InfiniBand}
	\label{perf_analysis_FDR}
\end{figure*}

An analysis of the application performance as measured by execution time is presented in this section. The cluster nodes in our experimental set up have 12 cores (up to 24 threads with hyper-threading) and therefore we use a maximum of 24 vGPUs (to avoid any noise due to CPU overhead). Up to 12 pGPUs will be used to map the vGPUs.

Figure~\ref{perf_analysis_QDR} and Figure~\ref{perf_analysis_FDR} show the time taken for data transfer and computation for varying pGPUs when the rCUDA framework is used over QDR and FDR InfiniBand. 
The `Overlapped data transfer and computation' label denotes that data transfers and computation are carried out concurrently on the same pGPU. 
The behaviour of the application is as expected. Multi-tenancy with sequential transfers allows for overlapping computations and data movement on the same pGPU, thus reducing the execution time. When QDR InfiniBand is used, time for data transfer without overlaps with communication is reduced up to 70\%, 84\%, 66\%, and 42\% when vGPUs are mapped to 1, 2, 4, and 6 pGPUs, respectively. In the case of FDR InfiniBand, the same time is 65\%, 77\%, 57\%, and 56\%. Consequently, the total power consumed is reduced but not indicated on the graph.

It is noted that when 12 pGPUs are used the data transfer times are not reduced further because (i) the execution time decreases with more pGPUs, and (ii) the data transfer time increases when more vGPUs are used allowing for little overlap between data transfers and computation on the same pGPU. This necessitates the need for determining the effective combination of pGPUs and vGPUs by estimating application perfomance both in terms of execution time and energy consumption. 

\subsection{Modelling Multi-tenancy for Performance and Energy Estimation}

An important challenge is to automatically determine the best multi-tenancy configuration for a deployment that can maximise performance (minimising execution time), but at the same time minimise the energy consumed. 

\subsubsection{Performance Model}
We firstly consider a basic model to account for execution time of the application when sequential data transfers are used with rCUDA, but without exploiting multi-tenancy. Subsequently, the model is optimised to take multi-tenancy into account. The model is then applied in the context of the hardware (NVIDIA Tesla K20 GPUs with QDR and FDR InfiniBand) we have employed in this research. 

The total execution time depends on: (i) time for transferring data and (ii) time for computing on the GPUs as shown in Equation~\ref{eq:total_exec_time}, which inherently depends on the number of GPUs (pGPUs or vGPUs) available to the application. 
%\begin{equation}
\begin{multline}
Total Execution Time = T_{transfer}(\#GPUs) + \\ T_{computation}(\#GPUs)
\label{eq:total_exec_time}
\end{multline}
%\end{equation}

Since there is perfect scalability for the computation times on the GPU (Section~\ref{cuda_scalability} and Section~\ref{rcuda_scalability}), the time required for computations by a given number of GPUs can be obtained as shown in Equation~\ref{eq:t_computation}. 
\begin{multline}
	T_{computation}(\#GPUs) = ComputationTime\_1pGPU ~ / ~ \\\#GPUs
	\label{eq:t_computation}
\end{multline}

The time to transfer the input data to all GPUs is shown in Equation~\ref{eq:t_transfer}. The time taken to allocate memory on each GPU using {\tt cudaMalloc()} and the time for moving small and large data structures to the GPUs are taken into account. Different data sizes achieve varying network bandwidth (Figure~\ref{figure5c}). To simplify the equation, the time to transfer data structures smaller than 100 bytes is denoted as $T_{small\_transfers}$\footnote{Data structures smaller than 100 bytes achieve the same bandwidth and are therefore grouped together. The InfiniBand frame size is typically 2~KB, which will be sent to the GPU in all cases where data is smaller than 100 bytes.}

\begin{multline}
	T_{transfer}(\#GPUs)	= \#GPUs~* (T_{cudaMalloc} +\\
							T_{small\_transfers}  
							+  T_{transfer\_4MB} + T_{transfer\_120MB})
							+\\ T_{transfer\_4GB}
	\label{eq:t_transfer}
\end{multline}

When multi-tenancy is taken into account there is an overlap between data transfers and computations on the same pGPU which reduces the total execution time. As shown in Figure~\ref{sequential_communications_2vgpu}, when 2 vGPUs are mapped onto a single pGPU, the time for data transfer is the time taken to move the first chunks of data to the pGPUs (until the completion of time step 12). The time for moving the remaining data chunks are not accounted for since it is overlapped by computation time. This is captured in Equation~\ref{eq:multitenancy1}.

\begin{multline}
Exec Time\_Multitenancy_{fully\_overlapped} =\\ 
T_{transfer}(\#vGPUs)~/~vGPUs\_per\_pGPU \\ 
+ vGPUs\_per\_pGPU * T_{computation}(\#vGPUs)
\label{eq:multitenancy1}
\end{multline}

If a very large number of vGPUs are used, then all data transfer times may not be overlapped with computation times. This can happen when the computation on the vGPU is not long enough to overlap data transfers to the pGPU and the computations on it. In this case, the total execution time depends on the time required to copy data to all the vGPUs and is shown in Equation~\ref{eq:multitenancy2}.

\begin{multline}
Exec Time\_Multitenancy_{not\_fully\_overlapped} =\\ T_{transfer}(\#vGPUs) + 
T_{computation}(\#vGPUs)
\label{eq:multitenancy2}
\end{multline}

As shown in Equation~\ref{eq:multitenancy} the maximum value from Equation~\ref{eq:multitenancy1} and Equation \ref{eq:multitenancy2} determines whether the application has significant overlaps between data transfer and computations.  

%Afterwards, the final formula for the theoretical model would be:

\begin{multline}
	Exec Time\_Multitenancy  = \\MAX (Exec Time\_Multitenancy_{fully\_overlapped},\\
	Exec Time\_Multitenancy_{not\_fully\_overlapped})
	\label{eq:multitenancy}
\end{multline}

Table~\ref{parameters_model_perf} shows actual values of the model for the experimental platform used in this research.

\begin{table}[t]
\caption{Time in seconds for GPU memory allocation and data transfer tasks of the financial risk application}
%\vspace{-0.25cm}
\label{parameters_model_perf}
\begin{center}
\begin{tabular}{ | l | c | c |}
\hline
  Parameter &   QDR &   FDR  \\ \hline \hline
$ComputationTime\_1pGPU$ & \multicolumn{2}{|c|}{9.55} \\ \hline
$T_{cudaMalloc}$ & 0.00267 &  0.0027 \\ \hline
$T_{small\_transfers}$ & 0.0048 & 0.0028 \\ \hline
$T_{transfer\_4MB}$ &  0.00133 &  0.00079 \\ \hline
$T_{transfer\_120MB}$ & 0.036   &  0.0205 \\ \hline
$T_{transfer\_4GB}$ & 1.171   &  0.67 \\ \hline
\end{tabular}
\end{center}
\end{table}

\begin{figure*}[t!]
\centering
\subfloat[1 to 4 pGPUs]
{\label{perf_model_QDR_1}
\includegraphics[width=0.48\textwidth]
{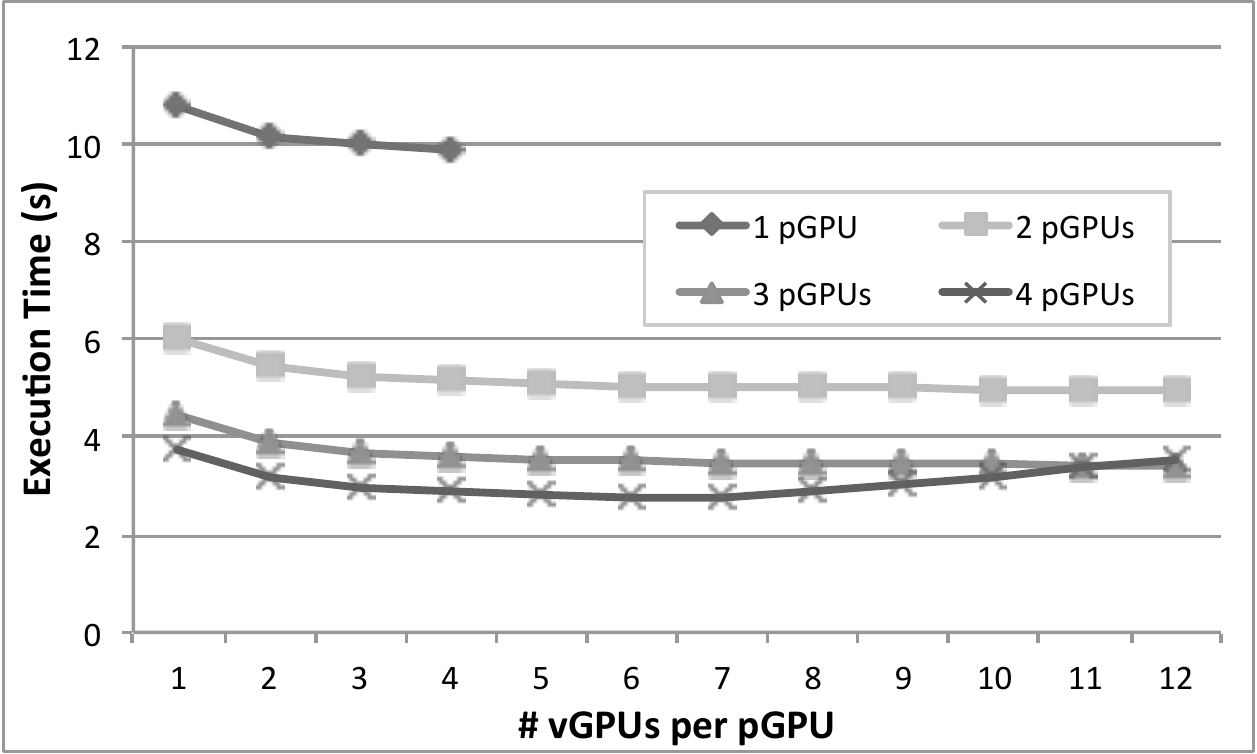}
}
\hfill
\subfloat[5 to 8 pGPUs]
{\label{perf_model_QDR_2}
\includegraphics[width=0.48\textwidth]
{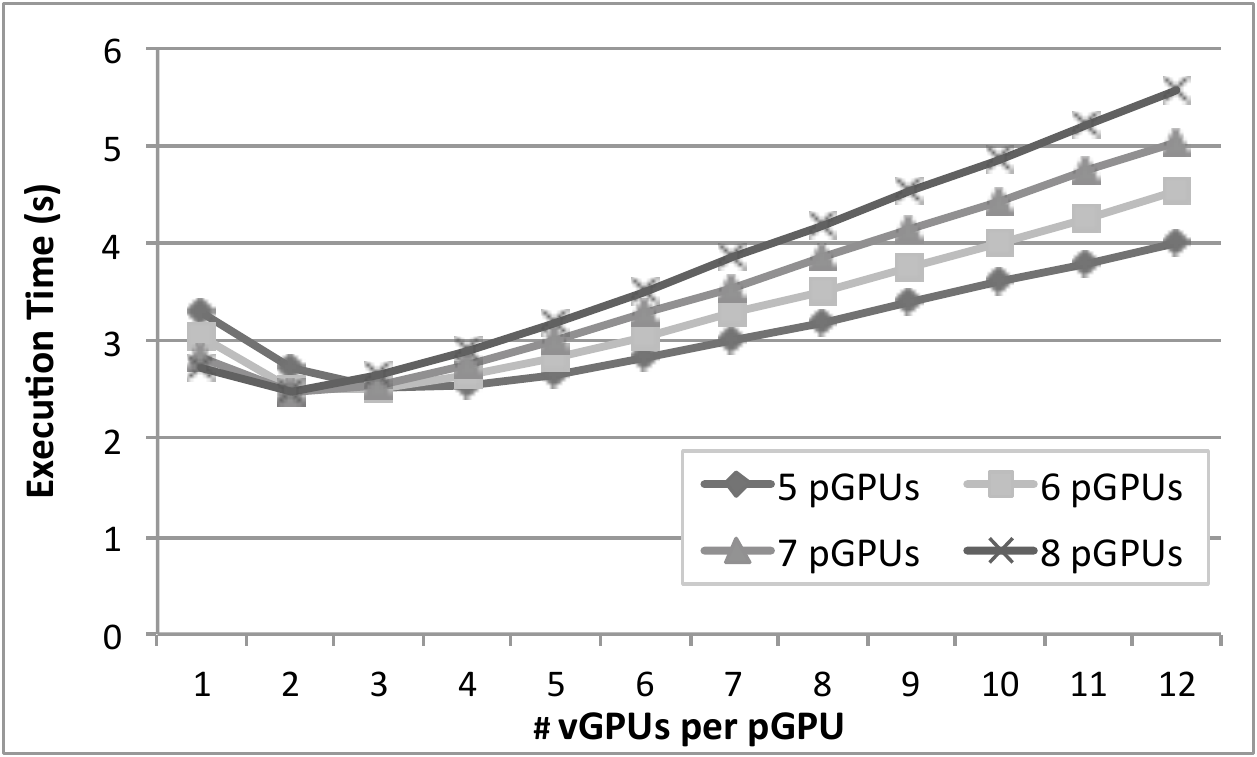} 
} \\

\subfloat[9 to 12 pGPUs]
{\label{perf_model_QDR_3}
\includegraphics[width=0.48\textwidth]
{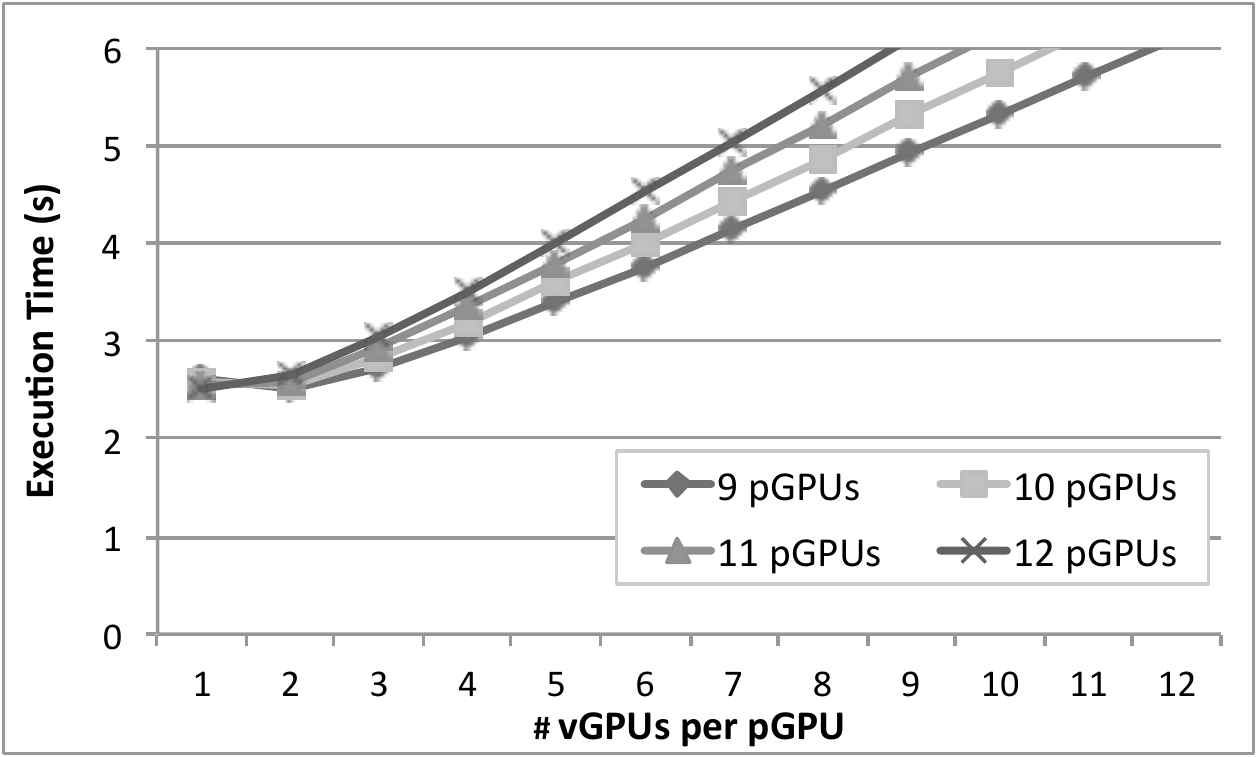}
}
\hfill
\subfloat[13 to 16 pGPUs]
{\label{perf_model_QDR_4}
\includegraphics[width=0.48\textwidth]
{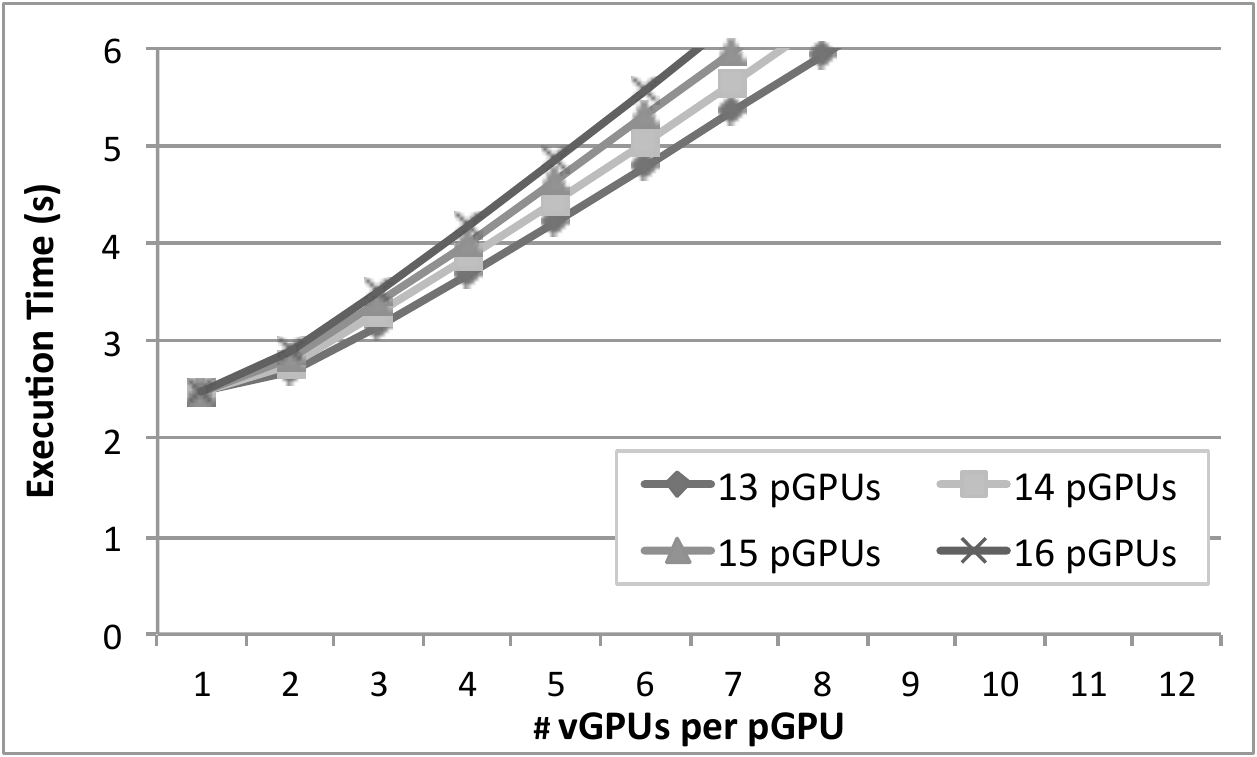} 
} \\

\caption{Results from performance model for QDR InfiniBand}
\label{perf_model_QDR}
\end{figure*}

\begin{figure*}[t!]
\centering
\subfloat[1 to 4 pGPUs]
{\label{perf_model_FDR_1}
\includegraphics[width=0.48\textwidth]
{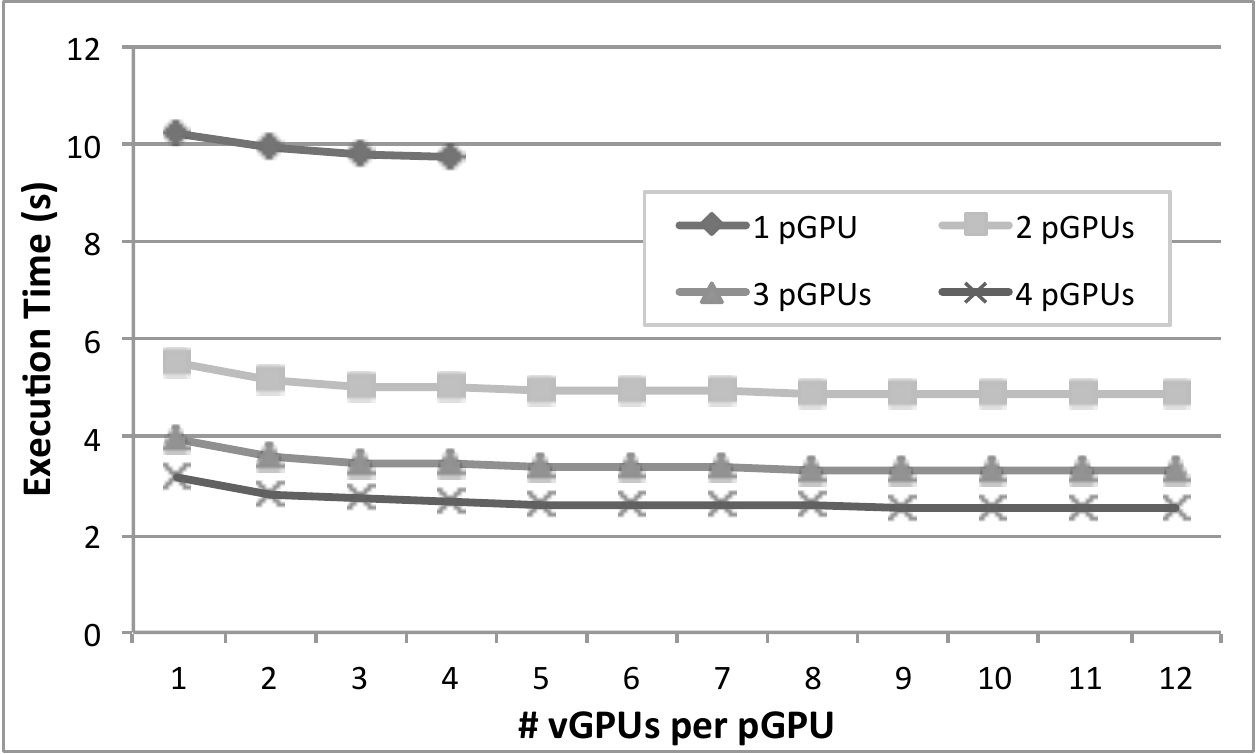}
}
\hfill
\subfloat[5 to 8 pGPUs]
{\label{perf_model_FDR_2}
\includegraphics[width=0.48\textwidth]
{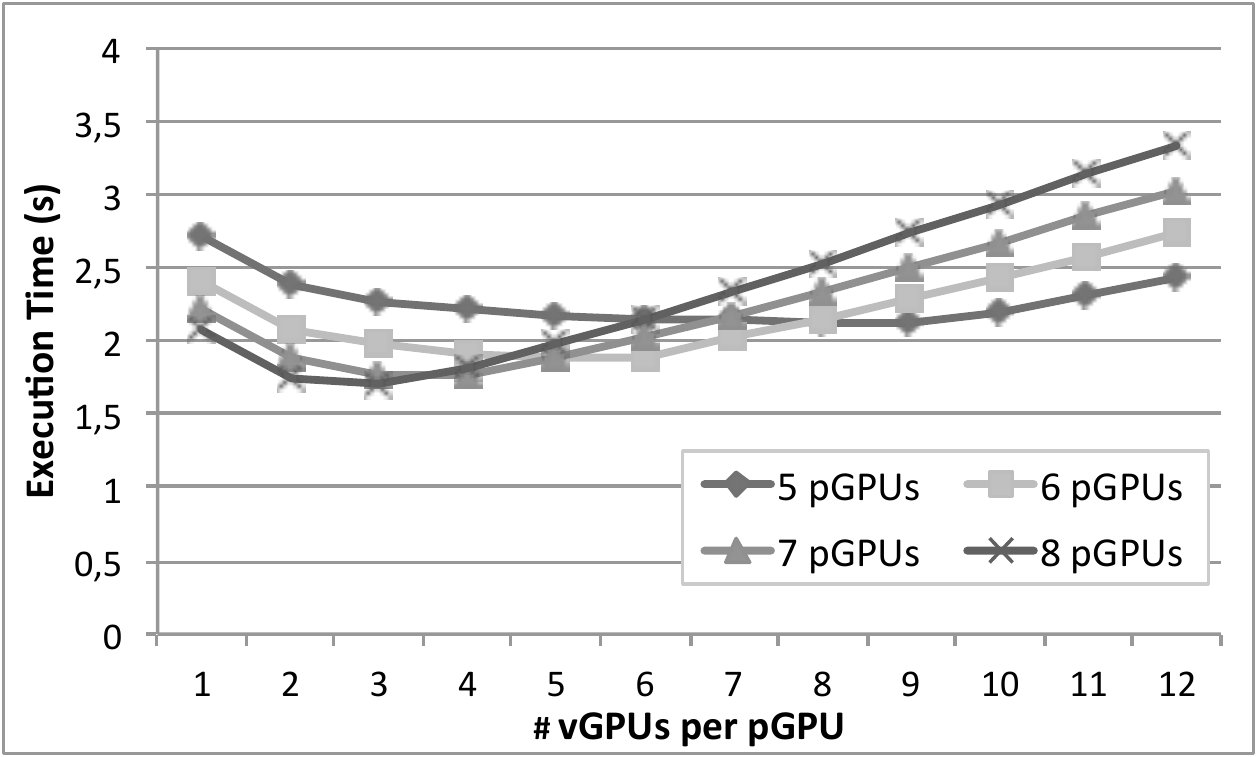} 
} \\

\subfloat[9 to 12 pGPUs]
{\label{perf_model_FDR_3}
\includegraphics[width=0.48\textwidth]
{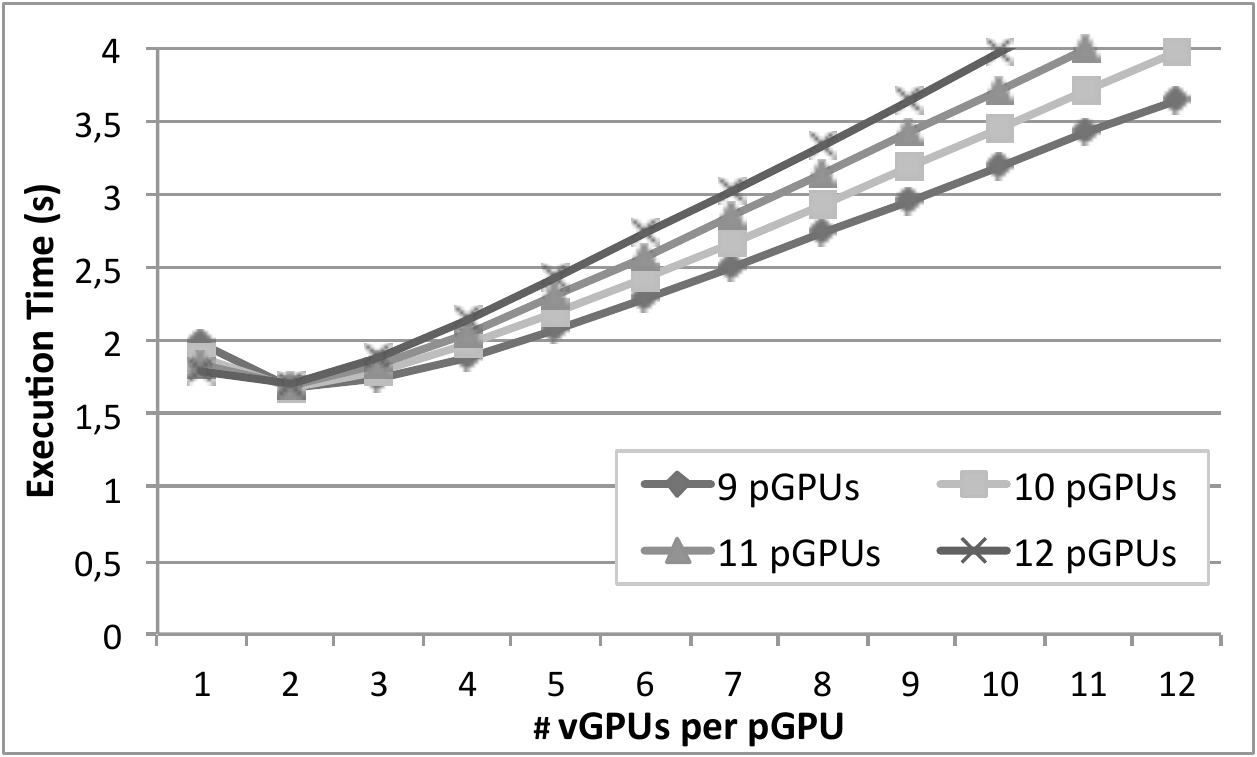}
}
\hfill
\subfloat[13 to 16 pGPUs]
{\label{perf_model_FDR_4}
\includegraphics[width=0.48\textwidth]
{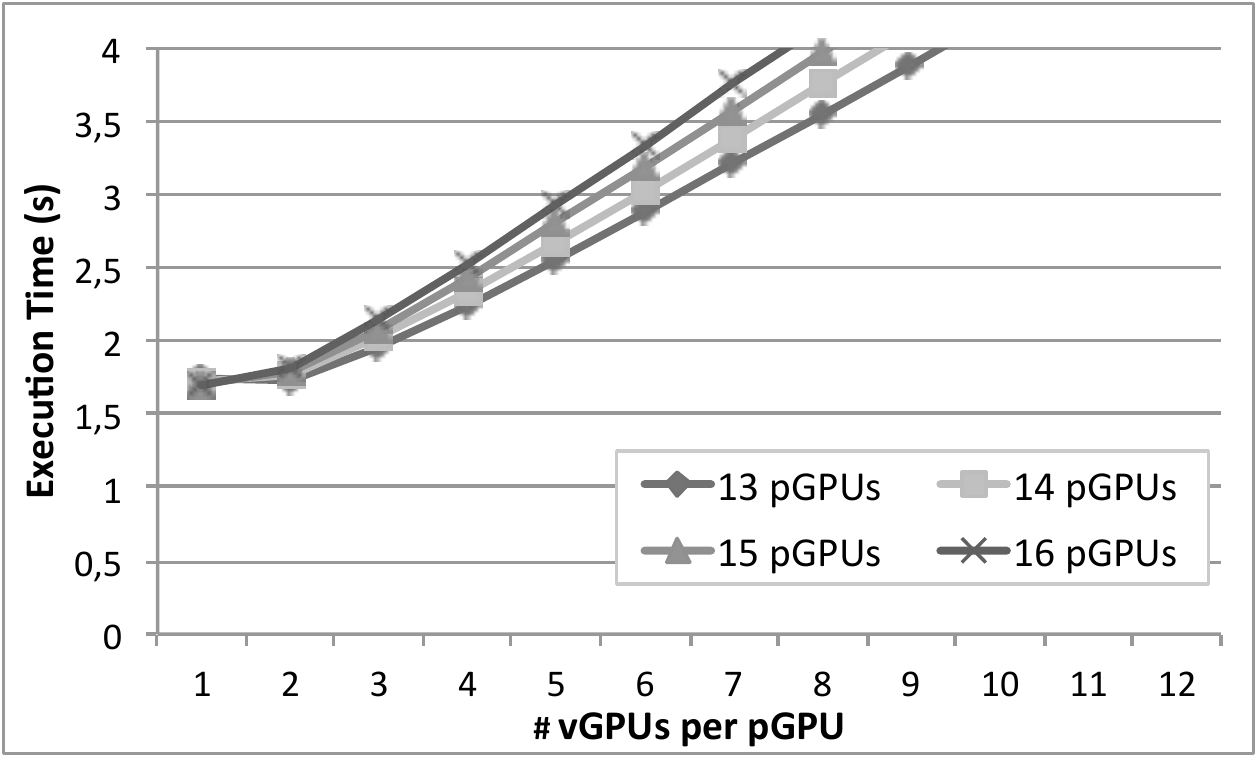} 
} \\

\caption{Results from performance model for FDR InfiniBand}
\label{perf_model_FDR}
\end{figure*}

Figure~\ref{perf_model_QDR} and Figure~\ref{perf_model_FDR} use these values in Equation~\ref{eq:multitenancy} for 1 to 16 pGPUs and up to 12 vGPUs per pGPU. The combinations of pGPUs and vGPUs that require the lowest execution time can be explored in this space. The estimated execution times are grouped for 1 to 4 pGPUs, 5 to 8 pGPUs, 9 to 12 pGPUs, and 13 to 16 pGPUs. In Figure~\ref{perf_model_QDR_1} and Figure~\ref{perf_model_FDR_1}, for one pGPU up to 4 vGPUs can be used. The total memory on the Tesla K20 devices is 4799~MB (from the {\tt nvidia-smi} command), which is exhausted by more than 4 vGPUs (total memory size consumed by the application on 4 vGPUs is 4484~MB). It is inferred from the figures that a large number of vGPU has detrimental effect on performance due to the overheads in data movements. Using QDR InfiniBand the model predicts a saturation sooner than FDR InfiniBand because of the overhead of data transfers due to a lower bandwidth available on the QDR network. The optimal deployment configuration of the application is 7 pGPUs with 2 vGPUs per pGPU and 9 pGPUs with 2 vGPUs per pGPU using QDR InfiniBand and FDR InfiniBand respectively. 

\subsubsection{Energy Model}

The amount of energy required to execute the application is modelled in this section. From Figure~\ref{sequential_communications} it is inferred that a GPU can be in the following four different states: (1) idle, (2) receive data, but no computations, (3) receive data and compute simultaneously, and (4) compute, but no data to receive.

Power is measured by querying {\tt nvidia-smi} every 200 milliseconds. The power required by the GPU in the first two states is the same. The NVIDIA Tesla K20 device requires 47 Watts while idling\footnote{The idle state in Figure~\ref{sequential_communications} is distinguished from the commonly known ``idle'' state. In Figure~\ref{sequential_communications}, the GPU has already been assigned to the application and therefore has been initialised by the GPU driver(this requires approximately 1.3 seconds in CUDA). After initialisation, the GPU does not perform any task, but actively waits for commands. In the commonly known ``idle'' state, the GPU is not assigned to an application and is not initialised by the driver. In this state, the Tesla K20 GPU requires 25 Watts.} and receiving data. The GPU requires 102 Watts in the last two states.

Using the above power readings for the four GPU states along with total 
execution time obtained from Equation~\ref{eq:multitenancy} an energy model is 
developed as shown in Equation~\ref{eq:energy}. The energy required by the GPU 
for computations (time spent on computations is obtained from 
Equation~\ref{eq:t_computation}) is eliminated to obtain the energy spent in the 
first and second states. The computation time on the pGPUs is 
$vGPUs\_per\_pGPUs~*~T_{computation}(\#vGPUs)$.

\begin{multline}
Total Energy  = \#pGPUs * ( T_{computation}(\#pGPUs)~ \\* ~102 ~ Watts ~ + 
(Exec Time\_Multitenancy - \\T_{computation}(\#pGPUs)) ~*~47 
~ Watts )
\label{eq:energy}
\end{multline}

\begin{figure*}[t!]
\centering
\subfloat[1 to 4 pGPUs]
{\label{ener_model_QDR_1}
\includegraphics[width=0.48\textwidth]
{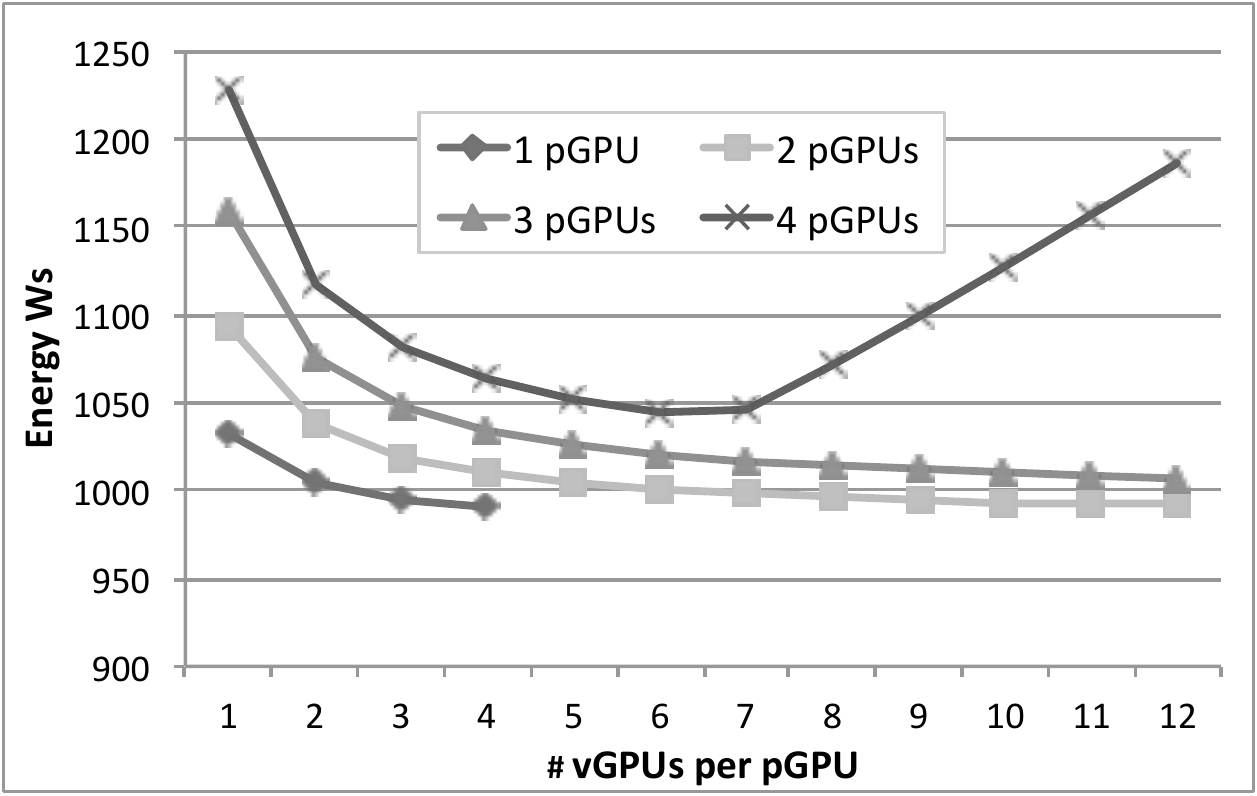}
}
\hfill
\subfloat[5 to 8 pGPUs]
{\label{ener_model_QDR_2}
\includegraphics[width=0.48\textwidth]
{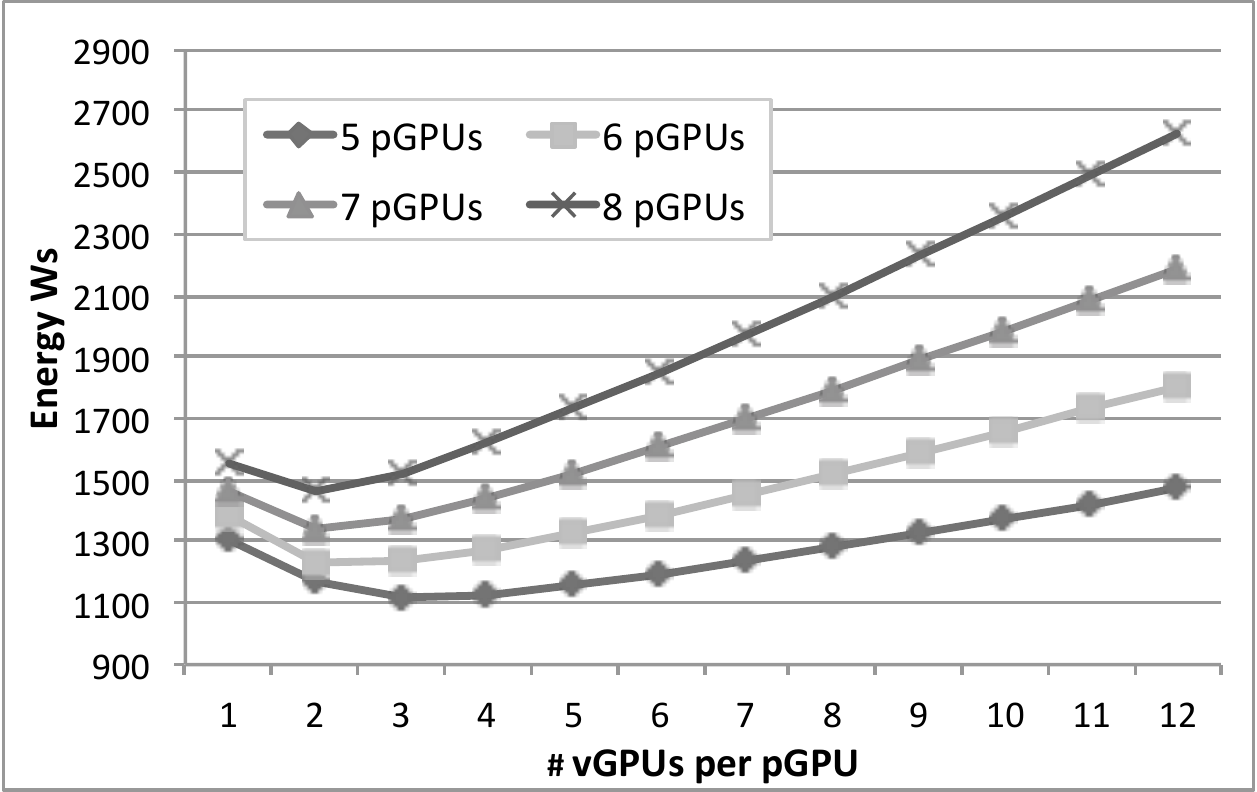} 
} \\

\subfloat[9 to 12 pGPUs]
{\label{ener_model_QDR_3}
\includegraphics[width=0.48\textwidth]
{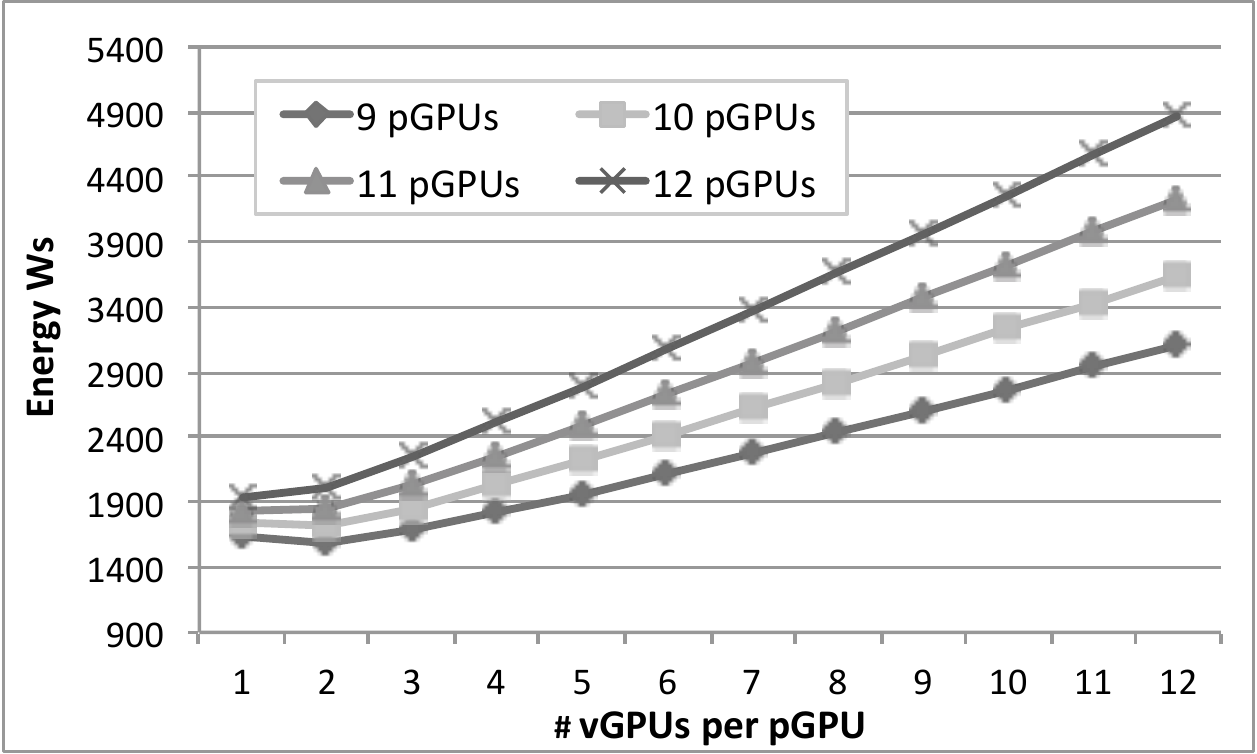}
}
\hfill
\subfloat[13 to 16 pGPUs]
{\label{ener_model_QDR_4}
\includegraphics[width=0.48\textwidth]
{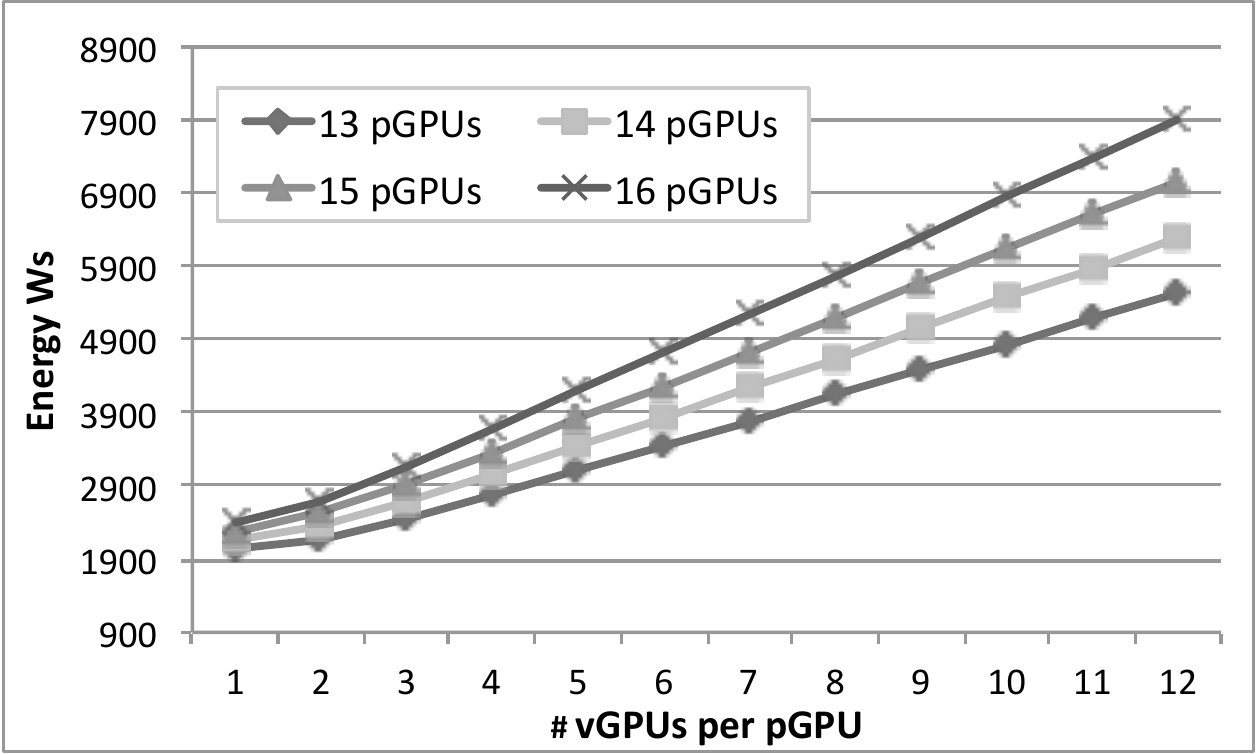} 
} \\

\caption{Results from energy model for QDR InfiniBand}
\label{ener_model_QDR}
\end{figure*}

\begin{figure*}[t!]
\centering
\subfloat[1 to 4 pGPUs]
{\label{ener_model_FDR_1}
\includegraphics[width=0.48\textwidth]
{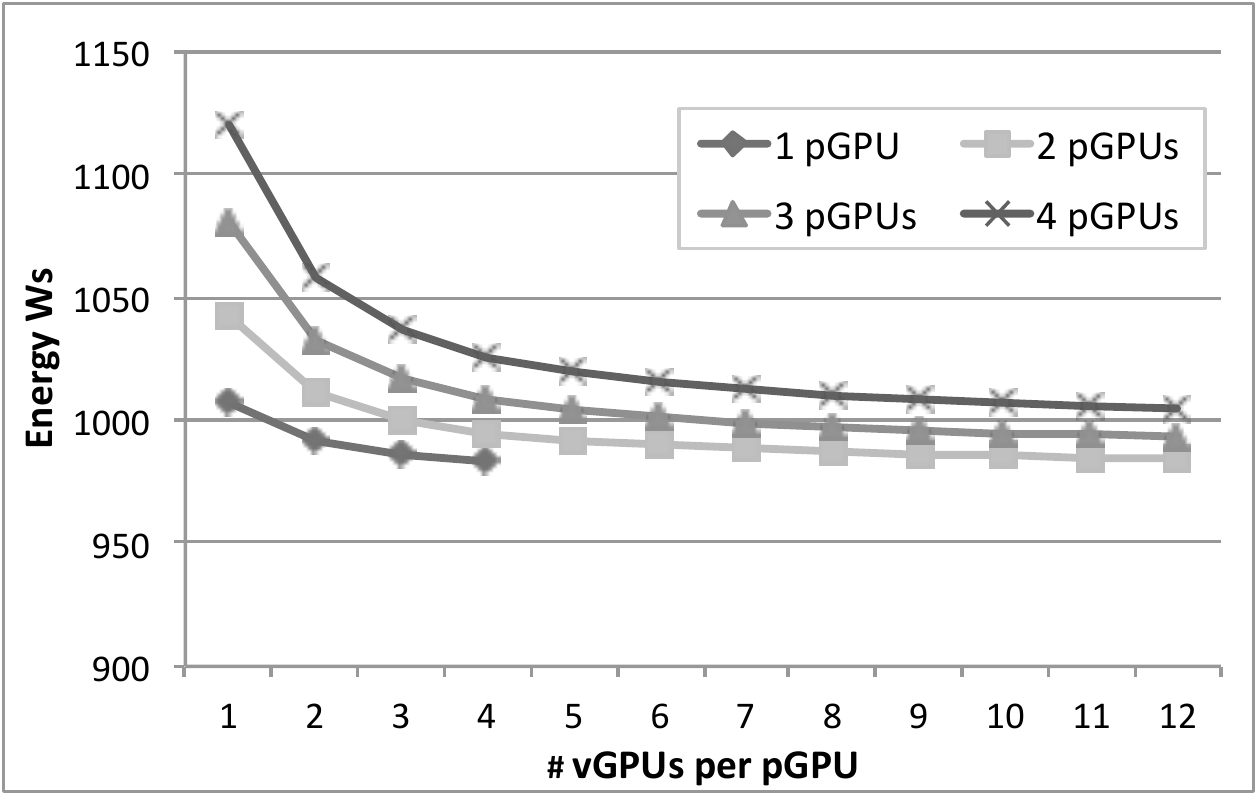}
}
\hfill
\subfloat[5 to 8 pGPUs]
{\label{ener_model_FDR_2}
\includegraphics[width=0.48\textwidth]
{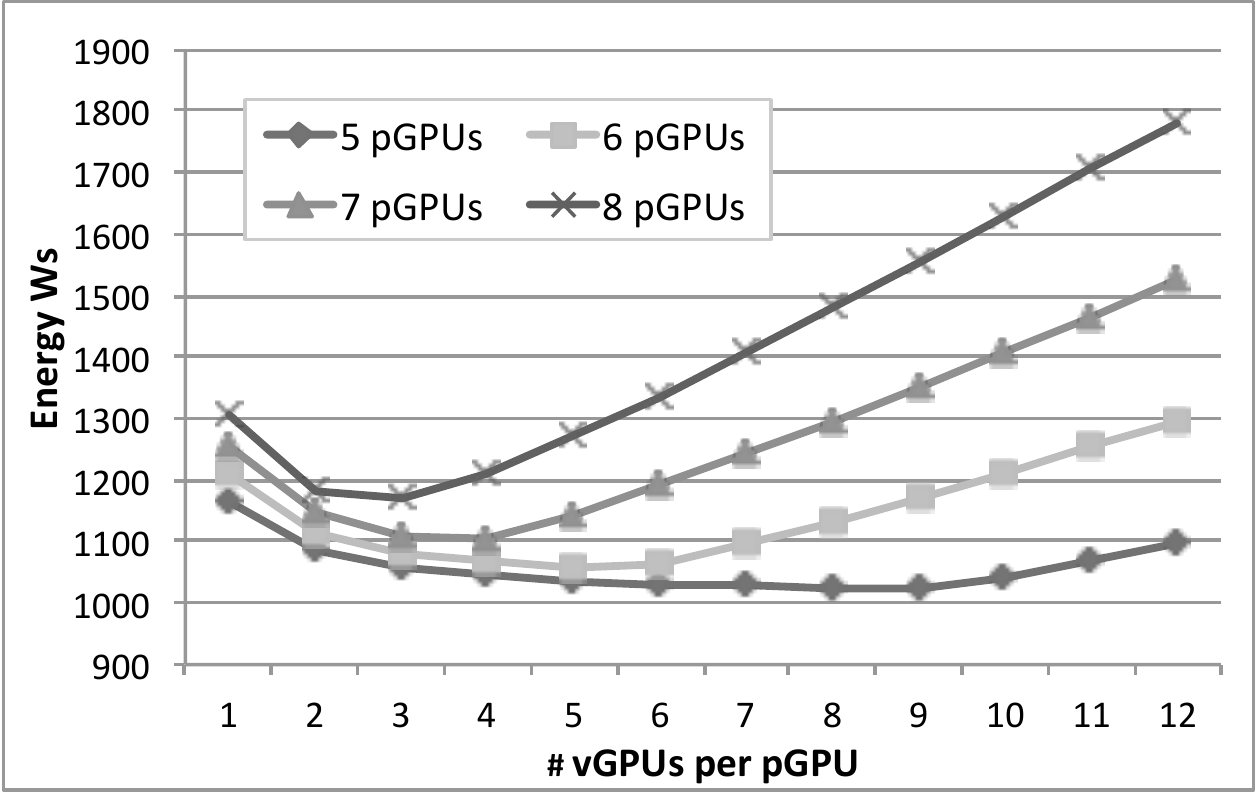} 
} \\

\subfloat[9 to 12 pGPUs]
{\label{ener_model_FDR_3}
\includegraphics[width=0.48\textwidth]
{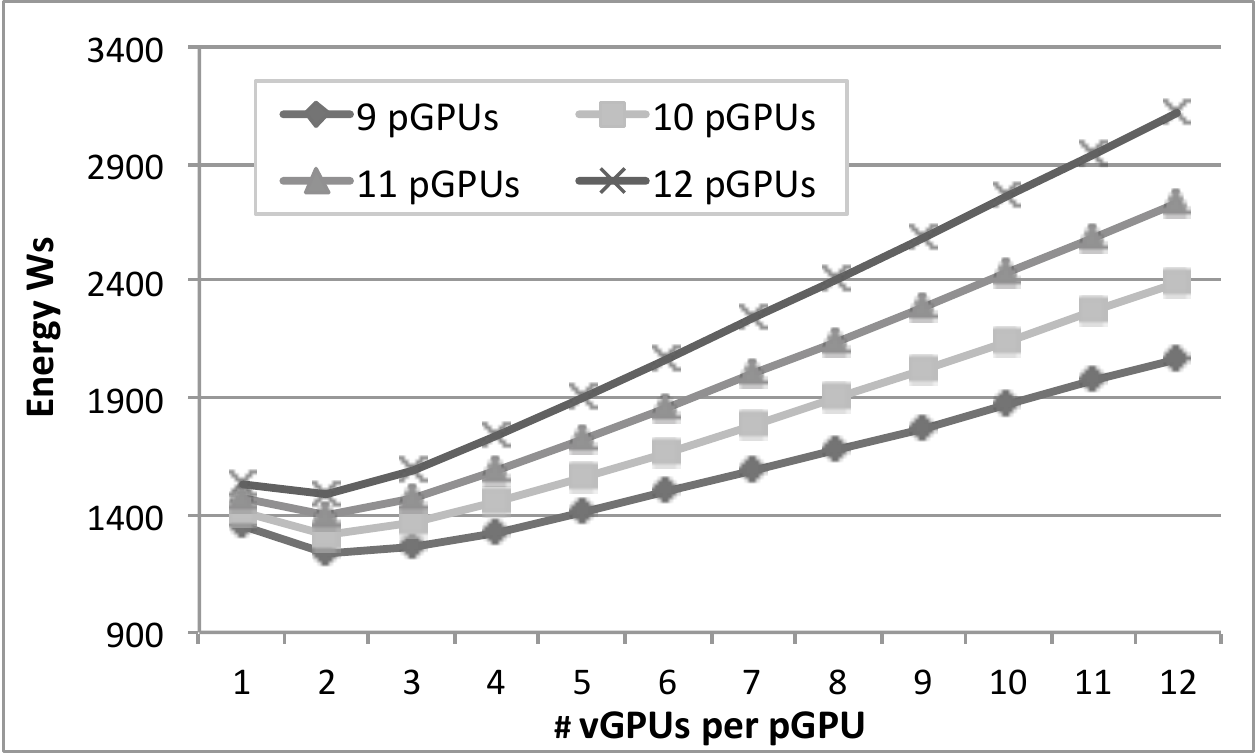}
}
\hfill
\subfloat[13 to 16 pGPUs]
{\label{ener_model_FDR_4}
\includegraphics[width=0.48\textwidth]
{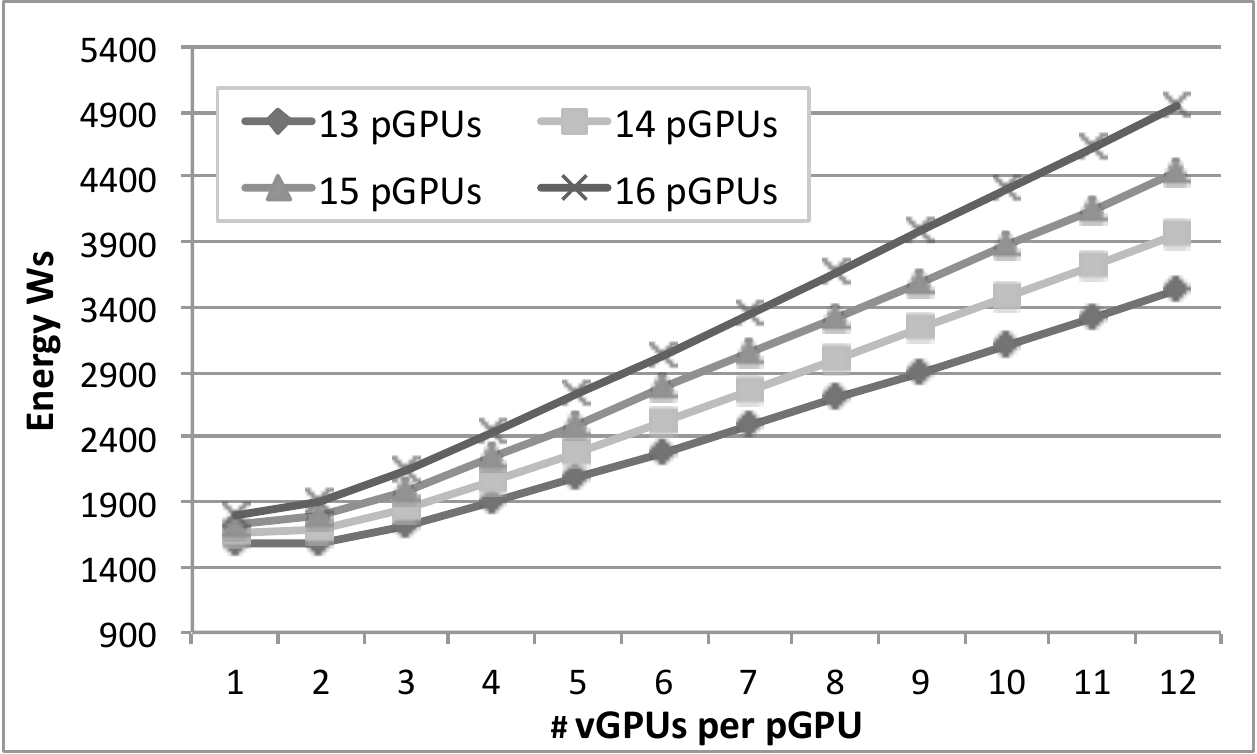} 
} \\

\caption{Results from energy model for FDR InfiniBand}
\label{ener_model_FDR}
\end{figure*}

\begin{figure*}[t!]
\centering
\subfloat[1 to 8 pGPUs]
{\label{combi_model_QDR_1}
\includegraphics[width=0.99\textwidth]
{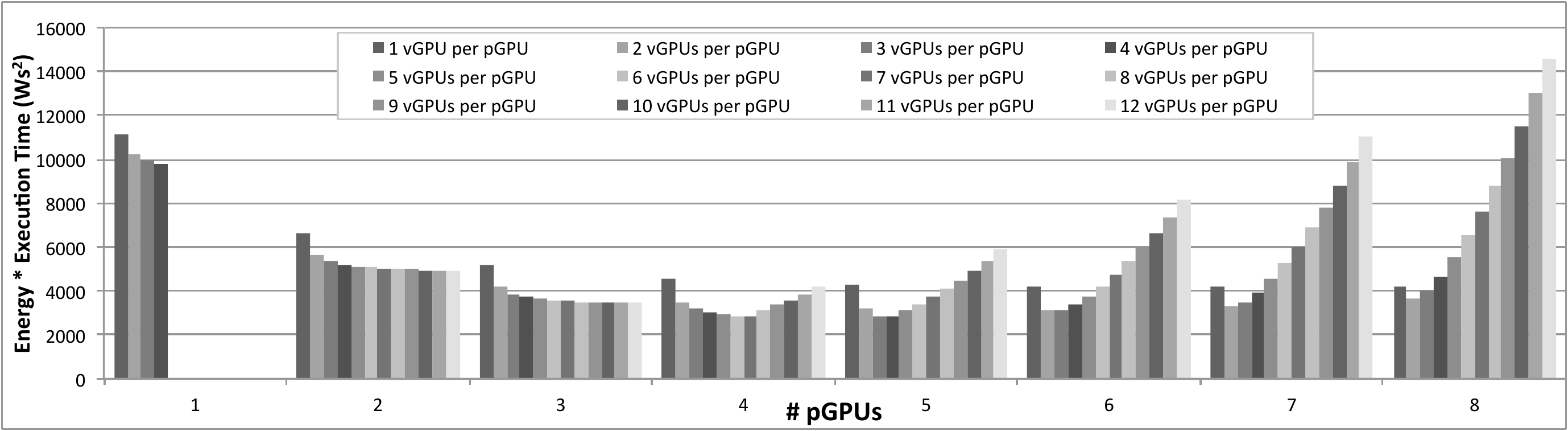}
}

\subfloat[9 to 16 pGPUs]
{\label{combi_model_QDR_2}
\includegraphics[width=0.99\textwidth]
{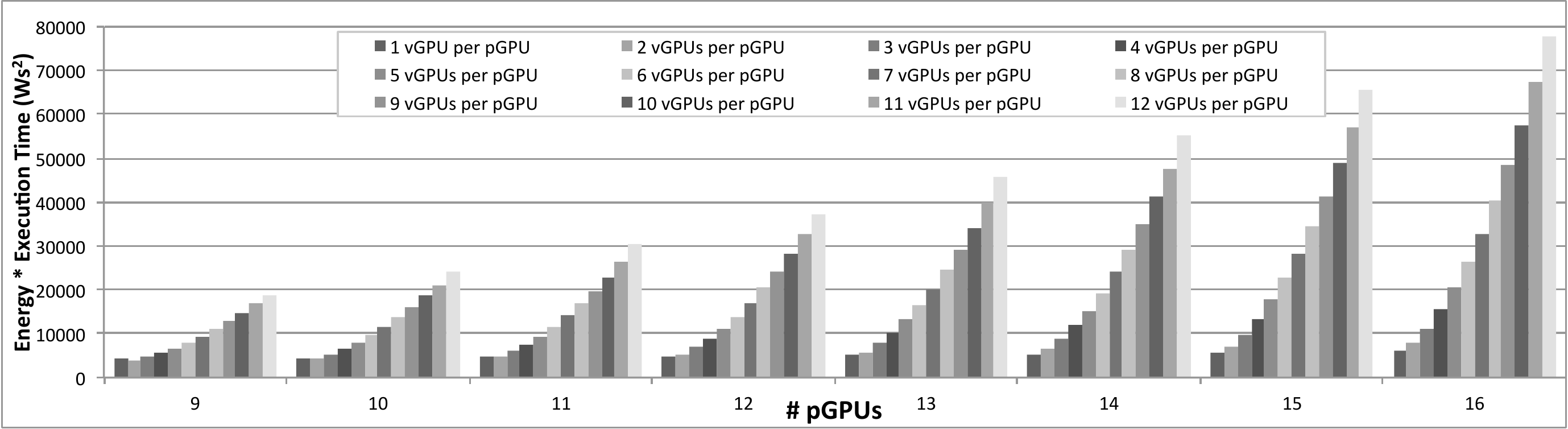}
}

\caption{Combined space of energy and execution time using QDR InfiniBand}
\label{combi_model_QDR}
\end{figure*}

\begin{figure*}[t!]
\centering
\subfloat[1 to 8 pGPUs]
{\label{combi_model_FDR_1}
\includegraphics[width=0.99\textwidth]
{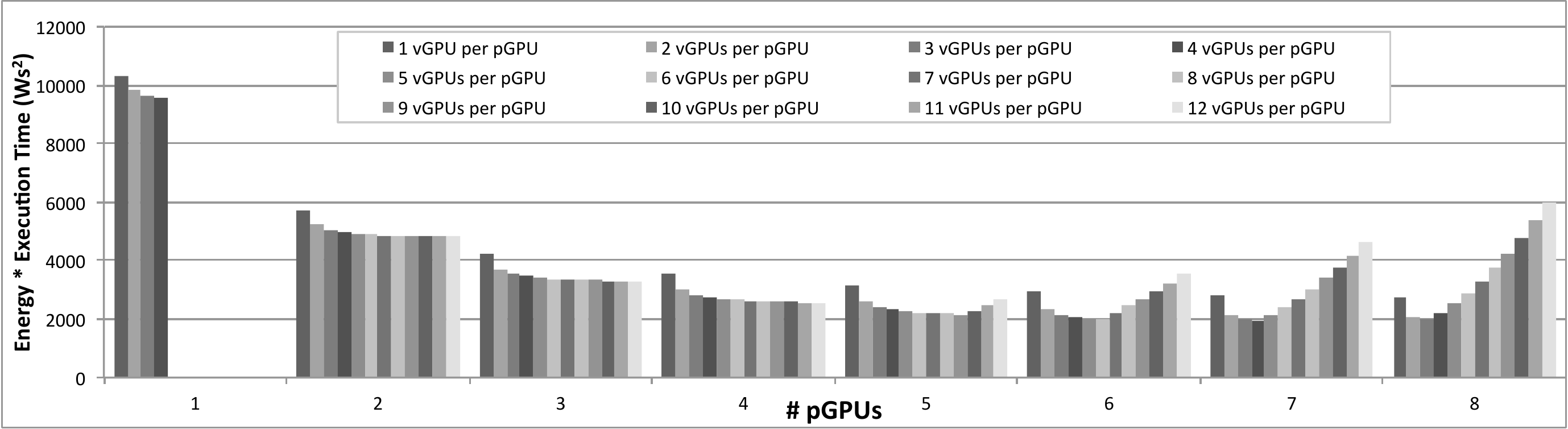}
}

\subfloat[9 to 16 pGPUs]
{\label{combi_model_FDR_2}
\includegraphics[width=0.99\textwidth]
{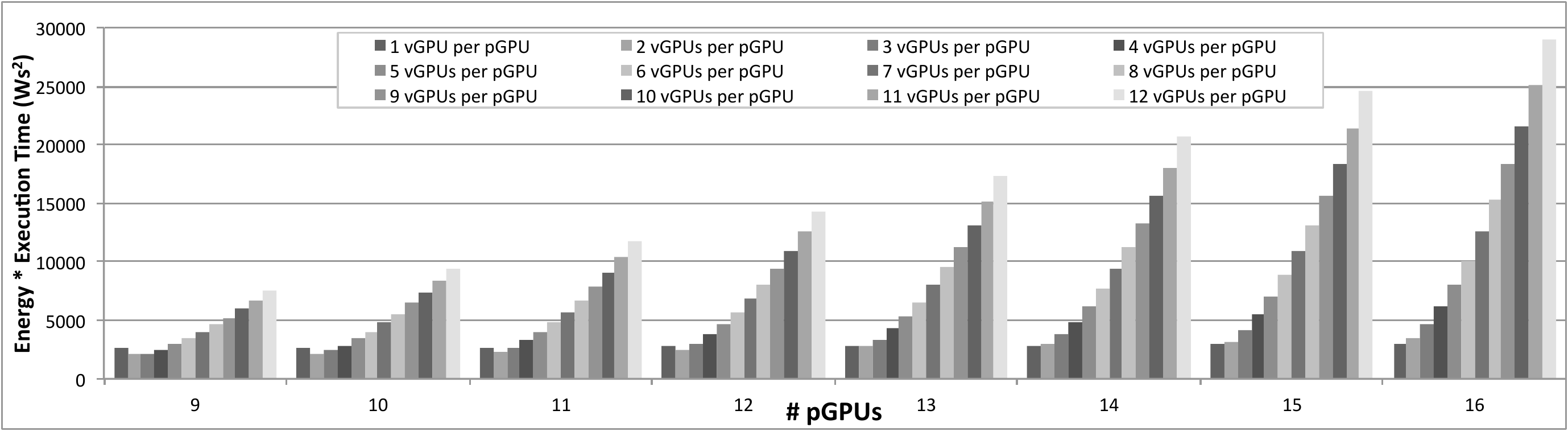}
}

\caption{Combined space of energy and execution time using FDR InfiniBand}
\label{combi_model_FDR}
\end{figure*}

Figure \ref{ener_model_QDR} and Figure~\ref{ener_model_FDR} present the results of the energy model from Equation~\ref{eq:energy}. It is noted that an energy efficient deployment is obtained using 4 vGPUs on 1 pGPU 
for both QDR InfiniBand and FDR InfiniBand. This is as expected given that the least amount of hardware is employed. However, there is a trade off since the lowest execution times are not obtained in this configuration. In Figure~\ref{combi_model_QDR} and Figure~\ref{combi_model_FDR}, an alternate space ($energy$ $*$ $execution$ $time$) is explored to find configurations that can maximise performance and minimise energy consumption.

\subsection{Generality of Proposed Approaches}
The financial risk application analysed in this paper is embarrassingly parallel and is representative of one class of workloads that execute in high-performance computing environments. In these workloads, the computations scale out linearly with the number of GPUs used. The research challenges which were initially posed are hence relevant to a wide range of accelerated applications that therefore can benefit from exploiting vGPUs, particularly in the context of multi-tenant vGPUs on a single pGPU. The approaches we have proposed as solutions to mitigate the challenges can thus be broadly applied to the benefit of these embarrassingly parallel applications. 

Typically, when accelerators are employed for embarrassingly parallel applications, the data necessary for computations will need to be transferred from the host to the memory of the used GPUs before computations can be actually performed. In the face of limited bandwidth for data transfers, linear scalability of the application will be affected degrading the overall performance of the application. However, by using our proposed approach of sequential data transfers along with the use of multi-tenancy on real GPUs, overall performance can be improved because data transfers from the host to the virtual GPUs can be overlapped with GPU computations for multiple physical GPUs. Such an approach effectively shares physical GPUs to optimise an application's execution time and energy consumption. 

There are multiple deployment options for an application when multi-tenancy is exploited, depending both on the characteristics of the application and those of the underlying hardware. Each application will have its own best combination of virtual GPUs that need to be mapped onto a physical GPU for best performance. Here our offline approach of modelling performance both in terms of energy and performance for estimations will be a handy method that can be broadly applied for other embarrassingly parallel applications.

Nevertheless, the approaches shown in the previous sections are less likely to hold for non-embarrassingly parallel applications. Each application may have its own challenges that will need to be addressed with unique approaches specific to each problem. However, given the complexity of modern day applications, it is more common to use specific approaches for a type of applications.

\section{Conclusions}
\label{conclusions}
In this paper, we have demonstrated the benefits of virtual GPUs for an application. Single tenancy (using one virtual GPU on a single physical GPU) and multi-tenancy (using a number of virtual GPUs on a physical GPU) were explored in this context. Concurrent and sequential data transfer models were considered. We hypothesised that multi-tenancy can improve the performance of the application. To validate the hypothesis the application was executed using rCUDA (remote CUDA), a framework that virtualises GPUs in a High-Performance Computing (HPC) cluster and provides remote GPUs to nodes that require acceleration on demand. Experimental results indicate that multi-tenant virtual GPUs with sequential data transfers optimise the performance of the application with less hardware when compared to single tenancy.

The research presented in this paper highlights that multi-tenant virtual GPUs can improve performance of an application. To achieve this we brought together the concepts of virtual GPUs and multi-tenancy in a single framework. The contribution of this research is to leverage multi-tenancy in the context of virtual GPUs within the rCUDA framework. Further, we have demonstrated this concept using a real world financial risk application of industrial use to optimise performance in terms of metrics, namely execution time, energy consumption and GPU utilisation. Given the application our research explores data transfer approaches with the aim of improving performance and how it is affected by memory and bandwidth bottlenecks. The experimental results provide insight that would not be apparent without a thorough evaluation. For example, it may be assumed that concurrent data transfers would improve performance, but the effect of memory and bandwidth limitations make sequential data transfers more appealing. The offline performance model is derived by making use of the experimental results which determines the configuration of the vGPU mapping on the pGPU for maximising performance.

%%Bibliography
\bibliographystyle{ieeetr}
\bibliography{references}

\end{document}